\documentclass[journal]{IEEEtran}
\usepackage{color}
\usepackage{import}
\usepackage{amsmath}
\usepackage{cite}
\usepackage{algorithmic}
\usepackage[ruled,vlined]{algorithm2e}
\usepackage{slashbox}
\usepackage{bm} 
\usepackage[pdftex]{graphicx}
\usepackage{amssymb}
\usepackage{multirow}
\usepackage{{physics}}
\usepackage[hidelinks,colorlinks=false]{hyperref}

\usepackage{xcolor}             

\newcommand{\lucas}[1]{\textcolor{blue}{#1}}

\usepackage{amsthm}  
\newtheorem{defin}{Definition}

\theoremstyle{definition}
\newtheorem{assump}{Assumption}
\newtheorem{remark}{Remark}
\newtheorem{example}{Example}

\newcounter{cons}
\newenvironment{cons}
{\par\noindent
	\refstepcounter{cons}%
	\textbf{Constraint C\thecons)}~\itshape\ignorespaces}
{\par\ignorespacesafterend}

\newcommand\ntpname{Note to Practitioners}
\newenvironment{ntp}
{\renewcommand\abstractname{\ntpname}\begin{abstract}}
{\end{abstract}}

\begin{document}
\title{Nonlinearity Compensation Based on Identified NARX Polynomials Models}

\author{Lucas~A.~Tavares, Petrus~E.~O.~G.~B.~Abreu and~Luis~A.~Aguirre
\thanks{Manuscript received November 24, 2020.}
\thanks{L. A. Tavares and P. E. O. G. B. Abreu are with the Graduate Program in
	Electrical Engineering, Universidade Federal de Minas Gerais, Belo Horizonte, MG, Brazil
	(e-mails: amarallucas@ufmg.br; petrusabreu@ufmg.br).}
\thanks{L. A. Aguirre is with the Department
	of Electronic Engineering, Universidade Federal de Minas Gerais, Belo Horizonte, MG, Brazil
	(e-mail: aguirre@ufmg.br).}}


\maketitle

\begin{abstract}

This paper deals with the compensation of nonlinearities in dynamical systems using nonlinear polynomial
autoregressive models with exogenous inputs (NARX). The compensation approach is formulated for static and
dynamical contexts, as well as its adaptation to hysteretic systems. In all of these scenarios, identified NARX models
are used. The core idea is to rewrite the model as an algebraic polynomial whose roots are potential compensation inputs.
A procedure is put forward to choose the most adequate root, in cases where more than one is possible.
Both numerical and experimental results are presented to illustrate
the method. In the experimental case the method is compared to other approaches. The results show that the proposed methodology can provide compensation input
signals that practically linearize the studied systems using simple and representative models with no more
than five terms.
\end{abstract}


\begin{ntp}
Nonlinearities impose significant performance limitations over a wide range of industrial applications, such as actuators and sensors. In many cases, classical control methods can present distress to deal with these effects. This paper is motivated by the use of compensation methods to improve the efficiency and quality of closed loop applications. Our goal is to present a practical technique based on NARX models for designing compensators. The method is tested with numerical examples, such as a model that describes a heating system with polynomial nonlinearity, and a hysteretic model to describe a piezoelectric actuator. An experimental pneumatic valve that presents several types of nonlinearities, including hysteresis or dead-zone, is used to illustrate the performance on a practical system.  As we propose a general method, the approach is also applicable to other systems modeled by NARX models.
\end{ntp}


%
\IEEEpeerreviewmaketitle
\vspace{-0.3cm}

\section{Introduction}

\IEEEPARstart{N}{onlinear} system identification is now a mature field. For the last three decades,
much attention has been paid to the problem of building nonlinear models from data
\cite{isermann2010identification,billings2013nonlinear,nelles2013nonlinear,schoukens2019}.
A natural next step is the use of such identified models in more specific applications, such as
compensation and control, which is the aim of this work.

The presence of nonlinear effects can impose significant performance limitations in a wide range of
applications, such as actuators \cite{Rakotondrabe2011,li2018position}, sensors \cite{cao2018temperature},
pH neutralization \cite{hong1996control}, and power systems \cite{singh2007neural}, among others
\cite{chernous2008control,morris2012,castillo2012soft}. A natural solution would be to mitigate such
effects by making the systems more linear -- by using a compensator -- and, consequently, more amenable to control. 
A dynamical model with simple structure is quite useful to design a compensator.

Compensation approaches aim to design a compensation input that reduces nonlinearity 
allowing for more accurate control and tracking. Many of these approaches start with an
appropriate model that represents the most fundamental aspects of the system, especially the static
nonlinearity \cite{agu_eal/02iee}. In the literature, there
is a vast number of works devoted to the modeling and compensation for nonlinear systems ranging from those based
on phenomenological models \cite{Rakotondrabe2011,Lin_etal2013, liu2011intrachannel} to those that use
computational intelligence \cite{Quaranta_etal2020} such as Radial Basis Functions (RBFs)
\cite{cao2018temperature,zhou2020dynamic,li2018position} and Neural Networks (NNs) \cite{singh2007neural,zhang2010neural,guo2018composite,meng2020neural},
among others.

The identification of phenomenological models tends to be a challenging task that requires sophisticated
algorithms based often on heuristics techniques. Therefore, satisfactory results depend largely on the proper design
of these algorithms, including the tuning of meta-parameters, which is usually an empiric task \cite{Quaranta_etal2020}.
In addition, the achievement of compensators for such models is not usually simple or even possible, because it depends
on the analytical inversion of these models. For methods based on universal approximation of functions, such as RBFs and NNs,
it is more difficult to provide a physics-based interpretation \cite{Quaranta_etal2020}. A particular type of neural network
that has been often used in the literature for identification and compensation of nonlinear systems is the Nonlinear
AutoRegressive with eXogenous inputs Neural Networks, i.e., NARX NNs \cite{zhang2010neural,meng2020neural}. 
Despite the benefits of NARX NNs due to the fair generality presented by NARX structures, these frameworks are 
based on a black-box philosophy that complicates the use of constraints related to the structure or parameters, 
which can be elegantly accommodated in gray-box approaches \cite{Abreu_etal2020}. Also, their compensators present low or no degree of interpretability which limits the analysis of these types of models and their compensators.

An alternative framework is based on NARX polynomial models, adopted in this paper. For this class of models,
if the structure is carefully chosen \cite{bil_eal/89,agu_bil/94b,pir_spi/03}, besides being quite general \cite{leo_bil/85a},
such models can encode nonlinear information in a simple and recognizable way \cite{agu_eal/02iee,Martins_Aguirre2016}, which
allows using them to derive explicitly compensation laws \cite{Abreu_etal2020}. In addition, NARX polynomials are amenable to
gray-box techniques \cite{Aguirre_2019} that allow the encoding relevant features from nonlinear systems, which is usually not
possible with purely black-box strategies. From now on, the term NARX models must be understood strictly as NARX polynomial
models, and form the basis of this work. Although these models can represent a variety of phenomena, few works
in the literature apply NARX polynomials for compensation since the most common applications take NARX NNs as a basis.

In the context of hysteresis compensation, \cite{Lacerda_etal2019} has presented a strategy based on an analytical
inversion of NARX models. For this purpose, somewhat restrictive assumptions must be satisfied by the model structure.
Also, as pointed out in \cite{Abreu_etal2020}, the
methodology developed by \cite{Lacerda_etal2019} may suffer from singularity problems when the velocity variable equals zero. 

Two ways to design compensators have been presented in \cite{Abreu_etal2020}:
the first one is similar to what was done in \cite{Lacerda_etal2019} and the second seeks compensators directly
from the data. Both strategies have overcome the singularity problem because the restrictions on the models' structure prevent a division by the velocity variable in the compensator. However, as the former also needs to isolate 
the input explicitly, such a method uses with more specific structures than those used in the present paper. 
The second strategy requires careful data processing, such as filtering the output signals. Also, some algebraic
tricks are required to overcome potential causality problems \cite{Abreu_etal2020}.

The main contributions of this work are the proposed approaches to find compensation inputs iteratively for nonlinear
systems in static and dynamical contexts through identified NARX models. Besides, an adaptation of the dynamical strategy
is presented for hysteretic systems. In both strategies, an algebraic polynomial of the compensation input is formulated,
which is achieved by manipulating the identified model. Thus, the compensation input signal is calculated iteratively,
which confers an adaptive feature to the approaches. The proposed compensators are compared with one well-established
\cite{Rakotondrabe2011} and two recent \cite{Abreu_etal2020} ones. The comparison is not performed with \cite{Lacerda_etal2019} due to the similarity with the first method of \cite{Abreu_etal2020}.

This work is organized as follows. In Section~\ref{back} background is provided. The statement of the compensation problem is
introduced in Sec.~\ref{sp}. Section \ref{Methodology} presents the compensation strategy proposed and formulated for static
(\ref{sub_comp_stat}), dynamical (\ref{sub_comp_dyn}) and adapted specifically for hysteresis (\ref{sub_comp_hys}) contexts. Numerical and experimental results are discussed in Sec.~\ref{NumericalExamples}. Finally, concluding remarks are given in Sec.~\ref{Conclusion}.
\vspace{-0.4cm}
\section{Background}\label{back}

A NARX (Nonlinear Autoregressive model with eXogenous inputs) polynomial model
${\cal M}$ for a single-input single-output system is given by \cite{leo_bil/85a}:
\begin{eqnarray}
\label{eq_model}
y(k) {=}  f^\ell\big(y(k{-}1), \ldots ,y(k{-}n_y),u(k{-}\tau_{\rm d}), \ldots ,u(k{-}n_u)\big) \nonumber \\
{+}e(k),
\end{eqnarray}

\noindent
where $u(k),\,y(k) \in \mathbb{R}$ are respectively the input and output signals sampled at instant
$k \in \mathbb{N}$, and $f^\ell(\cdot)$ is a nonlinear polynomial function with degree
$\ell \in \mathbb{N}^+$. $n_u,\,n_y \in \mathbb{N}^+$ are the maximum lags for $u$ and $y$,
respectively, $\tau_{\rm d} \in \mathbb{N}^+$ is the pure time delay, and $e(k)$ accounts for the
uncertainties and possible noise.

Model (\ref{eq_model}) is a parsimonious polynomial model in the sense that it contains
only a small group of regressors chosen from an usually large set of candidate regressors
by means of some structure selection procedure \cite{bil_eal/89,agu_bil/94b,pir_spi/03,fal_eal/15,ret_agu/19}.
Each regressor of ${\cal M}$, which can be any linear and nonlinear combination up to degree
$\ell$,  is multiplied by a constant parameter, indicated by $\theta_i$. Hence,
a NARX polynomial model is linear-in-the-parameters and classic least squares (LS) procedures
can be used \cite{Norton_1986}. In the presence of noise, however, it is common to add moving average (MA)
terms to the model, which will no longer be linear-in-the-parameters. Fortunately, extended least
squares estimators (ELS) can be used to circumvent noise-induced bias \cite{Ljung_1987,bil_eal/89}.

\subsection{Steady-state analysis}\label{sub_steady_state_analysis}

The steady-state relation of model (\ref{eq_model}) is obtained by taking
$u(k)=\bar{u}$ and $y(k)=\bar{y}, ~\forall k$, thus yielding:
\begin{equation}
\bar{\cal M}: ~ \bar{y} =  \bar{f}^{\ell}(\bar{u},\bar{y}),
\nonumber
\end{equation}

\noindent
which, {\it for a known value of $\bar{u}$}, can be rewritten as:
\begin{equation}
\label{eq_model_stat}
c_{y,\,\ell_y}(\bar{u})\bar{y}^{\ell_y}+c_{y,\,\ell_y{-}1}(\bar{u})\bar{y}^{\ell_y{-}1}+\ldots+c_{y,1}(\bar{u})\bar{y}+ c_{y,0}(\bar{u}) = 0, 
\end{equation}

\noindent
where $1 \leq \ell_y \leq \ell$ is the degree of the static model $\bar{\cal M}$, whose coefficients
$c_{y,i},~i=0,\ldots,\ell_y$ usually depend on $\bar{u}$. Solving (\ref{eq_model_stat}) for the unknown
$\bar{y}$ is achieved by finding the $\ell_y$ roots of this polynomial. The roots of (\ref{eq_model_stat})
will yield the fixed points of model (\ref{eq_model}) for $\bar{u}$, whose definition is presented below.

\begin{defin}{\rm(Fixed points \cite{Aguirre_Mendes1996})}.
	\label{Definition:Fixed_Points}
	The steady-state analysis of model (\ref{eq_model}) is computed by taking $y(k){=}\bar{y},\,\forall k$
	and $u(k){=}\bar{u},\,\forall k$, yielding $\bar{y} = \bar{f}^{\ell}(\bar{y},\bar{u})$, whose
	solution/root(s) $\bar{y}$ (\ref{eq_model_stat}) for a given constant value of input $\bar{u}$
	is defined as the fixed point(s), or equilibria, of model (\ref{eq_model}) for $\bar{u}$.
\end{defin}

The condition for (local) stability of the fixed points is obtained by finding the eigenvalues
of the Jacobian matrix of model $\cal M$ (\ref{eq_model}) evaluated at each fixed point, as follows:
\begin{equation}
\label{eq_stab_fix_point}
\bigg|{\rm{eig}} \left(\frac{\partial f^{\ell}}{\partial \bm{y}} \Big|_{\bar{u},\bar{y}} \right) \bigg| < 1, 
\end{equation}

\noindent
where $\bm{y}=[y(k-1)\, \ldots y(k-n_y)]^T$, $T$ is the transpose and ${\rm eig}(\cdot)$ indicates the
eigenvalues.

\begin{example}\label{ex_1}
	Consider model ${\cal M}$ given by:
	\begin{eqnarray}
	\label{Eq:Example_Model_Stat}
	y(k) & {=} & \hat{\theta}_1y(k-1)+\hat{\theta}_2u(k-1)+\hat{\theta}_3u(k-1)u(k-2) \nonumber \\
	     & {~} & +\hat{\theta}_4u(k-1)^2+\hat{\theta}_5u(k-1)^3,
	\end{eqnarray}
	
	\noindent
	for which $\tau_{\rm d} {=} 1$, $n_y{=}1$, $n_u{=}2$, and $\ell{=}3$. Its static form $\bar{\cal M}$ is obtained taking
	$u(k-1){=}u(k-2){=}\bar{u}$ and $y(k-1){=}y(k){=}\bar{y}$, such that:
	\begin{equation}\label{eq_model_ss}
	\bar{y}  =   \hat{\theta}_1\bar{y} +\hat{\theta}_2 \bar{u}+\hat{\theta}_3\bar{u}^2 +\hat{\theta}_4\bar{u}^2+\hat{\theta}_5\bar{u}^3,
	\end{equation}
		
	\noindent
	which can be written in the format of (\ref{eq_model_stat}) as:		
	\begin{equation}
		0  =   \underbrace{\big[\hat{\theta}_1-1\big]}_{c_{y,1}}\bar{y} + \underbrace{\hat{\theta}_5\bar{u}^3 +\big[\hat{\theta}_3+\hat{\theta}_4\big]\bar{u}^2  + \hat{\theta}_2 \bar{u}}_{c_{y,0}(\bar{u})}.
	\end{equation}
	
	\noindent
	Hence, model (\ref{Eq:Example_Model_Stat}) only has one fixed point for each value of $\bar{u}$, given by:
	\begin{equation}
		\label{fp1}
		\bar{y} = -\dfrac{c_{y,0}(\bar{u})}{c_{y,1}} = \dfrac{\hat{\theta}_5\bar{u}^3+\big[\hat{\theta}_3+\hat{\theta}_4\big]\bar{u}^2+\hat{\theta}_2\bar{u}}{1-\hat{\theta}_1}.
	\end{equation}
	
	\noindent
	For the first-order model (\ref{Eq:Example_Model_Stat}),
	the Jacobian ``matrix'' will be a scalar and condition (\ref{eq_stab_fix_point}) 
	becomes:
	\begin{eqnarray}
	\label{Eq:example1}
	\left| \frac{\partial f^{\ell}}{\partial y(k-1)} \Big|_{\bar{u},\bar{y}} \right| & < & 1, \nonumber \\
	\left|\hat{\theta}_1 \Big|_{\bar{u},\bar{y}} \right| & < & 1, \nonumber \\
	-1 ~<~ \hat{\theta}_1 & < & 1.
	\end{eqnarray}
	
	\noindent
	Therefore, if (\ref{Eq:example1}) is satisfied, then (\ref{fp1}) is a stable fixed point. 
\flushright{$\square$}

\end{example}

\section{Statement of the Problem}
\label{sp}

\begin{figure}[h]
	\centering{
		\centering
		\includegraphics[width=1\columnwidth]{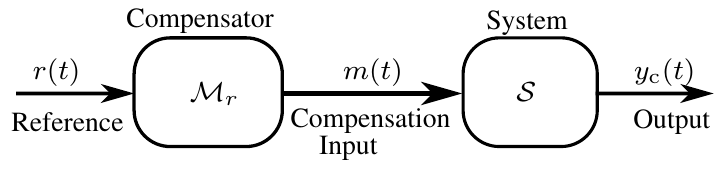}
		\vspace{-0.8cm}		
		\caption{Block diagram of the compensated system.}
		\label{block_diagram}
	}
\end{figure}

It is assumed that a NARX model $\cal{M}$ (\ref{eq_model}) is available for a
given nonlinear dynamical system $\cal{S}$, estimated
from input-output data $Z^N=\{u(k),\,y_{\rm s}(k)\}_{k=1}^N$ collected from $\cal{S}$.
Based on ${\cal M}$, the aim is to design a compensator ${\cal M}_r$  such that 
the open-loop combination of ${\cal M}_r$ followed by 
$\cal{S}$ (see Fig.\,\ref{block_diagram}) is more linear and therefore more amenable
for control. Specifically, ${\cal M}_r$ should compensate the nonlinearity in $\cal{S}$.

Mathematically, we require that the reference $r(k)$ and output $y_{\rm c}(k)$ of the
compensated system should be closer than the input $u(k)$ and output $y_{\rm s}(k)$
of the uncompensated system. Hence $J[r(k),\,y_{\rm c}(k)]<J[u(k),\,y_{\rm s}(k)]$,
where $J$ is some proximity cost function, like the mean squared value.

\section{Methodology}
\label{Methodology}

This section presents the methodology developed to design compensators based on
NARX polynomial models. First, in Sec.\,\ref{sub_comp_stat}, we present the static 
compensation that is simpler to understand and serves as a basis for the main result,
which is the dynamical compensation detailed in Sec.\,\ref{sub_comp_dyn}. In the sequel,
this approach is applied to systems with hysteresis in Sec.\,\ref{sub_comp_hys}.
The identification of model ${\cal M}$ is not described in this paper. The interested
reader is referred to \cite{Aguirre_2019}, and references therein.

\subsection{Static Compensation}
\label{sub_comp_stat}

It is assumed that the input and output signals are bounded, thus
$u(k) \in [u_{\rm min},\, u_{\rm max}], \forall k$ and $y_{\rm s}(k) \in [y_{{\rm min}},\, y_{{\rm max}}], \forall k$.
As for model $\cal{M}$(\ref{eq_model}), the following assumptions will be needed:

\begin{assump}
	\label{Assump1}
	$\cal{M}$ is valid, that is, $y(k) \approx y_{\rm s}(k)$ for the same input.
\end{assump}

\begin{assump}
	\label{Assump2}
	For any $\bar{u} \in [u_{\rm min},u_{\rm max}]$, the model
	${\cal M}$ has at least one local stable fixed point such that
	$\bar{y} \in [y_{\rm min}, y_{\rm max}]$.
\end{assump}

In order to obtain a static compensator $\bar{\cal M}_r$, an {\it inverse problem} in steady-state has to be solved.
In other words, we seek the system input  values $\bar{m}$ that will drive the output to the reference at steady-state,
$\bar{y} \approx \bar{r}$. Considering the model $\cal{M}$ expressed as $\bar{y} = \bar{f}^{\ell}(\bar{u},\bar{y})$,
$\bar{y}$ is replaced by $\bar{r}$ and $\bar{u}$ by $\bar{m}$ so that $\bar{\cal M}_r$ is $\bar{r} = \bar{f}^{\ell}(\bar{m},\bar{r})$.
Consequently, it is possible to rewrite (\ref{eq_model_stat}) by grouping its terms in such a way that it yields a
polynomial in the unknown variable $\bar{m}$, $\cal {\bar M}_{\rm r}$:
\begin{eqnarray}
\label{eq_comp_stat}
c_{m,\,\ell_m}({\bar{r}}){\bar{m}}^{\ell_m} {+} c_{m,\,\ell_m{-}1}({\bar{r}}){\bar{m}}^{\ell_m{-}1}{+} \ldots{+}c_{m,\,1}({\bar{r}})\bar{m}\nonumber \\
{+} c_{m,\,0}({\bar{r}}) {=} 0.
\end{eqnarray}

Although each of the $\ell_m$ roots of (\ref{eq_comp_stat}) is a solution to the inverse problem,
not all are appropriate to be used in practice. For this  reason, two constraints are considered. The used root:

\begin{cons}
	\label{c1}
	must be real, ${m} \in \mathbb{R}$; and
\end{cons}

\begin{cons}
	\label{c2}
	must be within the data range, namely ${m} \in [u_{\rm min},\, u_{\rm max}]$.
\end{cons}

Assumption\,\ref{Assump2} ensures that any root $\bar{m}$ (\ref{eq_comp_stat}) that satisfies the above
constrains will drive the system to a stable fixed point.
Also, because of Assumption~\ref{Assump1}, such a steady-state will satisfy $\bar{y}_{\rm c} \approx {\bar{r}}$.
The algebraic procedure is illustrated below with a simple example. Then, in the sequel, a dynamical version of
this procedure will be discussed in Sec.\,\ref{sub_comp_dyn}.

\begin{example}\label{ex_3}
	For the model in Example\,\ref{ex_1}, it was seen that:
	\begin{equation*}
	\bar{y}  =   \hat{\theta}_1\bar{y} +\hat{\theta}_2 \bar{u}+\hat{\theta}_3\bar{u}^2 +\hat{\theta}_4\bar{u}^2+\hat{\theta}_5\bar{u}^3,
	\end{equation*}
	
	\noindent
	which can be written in the format of (\ref{eq_comp_stat}) as:		
	\begin{eqnarray}
	\label{pol_ex3}
	\!\!\!\!\!\!\bar{r} & {=} &  \hat{\theta}_1\bar{r} +\hat{\theta}_2 \bar{m}+\hat{\theta}_3\bar{m}^2 +\hat{\theta}_4\bar{m}^2+\hat{\theta}_5\bar{m}^3, \nonumber \\
	\!\!\!\!\!\!0 & {=} &  \underbrace{\hat{\theta}_5}_{c_{m,3}}\bar{m}^3 +\underbrace{\big[\hat{\theta}_3+\hat{\theta}_4\big]}_{c_{m,2}}\bar{m}^2  + \underbrace{\hat{\theta}_2}_{c_{m,1}}\bar{m} +\underbrace{\big[\hat{\theta}_1\bar{r}-\bar{r}\big]}_{c_{m,0}(\bar{r})}.
	\end{eqnarray}
	
	\noindent
	Hence, for a given reference value $\bar{r}$, the roots of (\ref{pol_ex3})	provide potential
	compensation inputs that in steady-state would drive the system to the target. A practical problem
	is to decide which of the three roots in this example should be used. If there is only one real root,
	then it is chosen as the compensation input. However, if there are three real roots, a more general
	decision-making process is required. \hfill $\square$
	
\end{example}

For the sake of clarity, the roots of \eqref{eq_comp_stat} will be placed in a 
vector $\bm{\bar{m}}(\bar{r}) \triangleq [\bar{m}_{1}~\cdots~\bar{m}_{\ell_m}]^T$. 
Although the $\ell_m$  values of $\bar{m}$ are solutions to the
inverse problem, only those that satisfy C\ref{c1} and C\ref{c2} should be considered as potential
compensation inputs. This reasoning underlies the main algorithm that will be used in the dynamical
context. The practical issue of {\it how}\, to choose from two potential inputs that satisfy
C\ref{c1} and C\ref{c2} will be discussed in the dynamical setting below.

\subsection{Dynamical Compensation}
\label{sub_comp_dyn}

The main difference between the framework developed in this subsection and the basis laid down
in the previous one is that here the reference is a sequence of values $r(k)$, and not a
constant value $\bar{r}$. The same is true for the compensation input $m(k)$ and the compensated
output $y_{\rm c}(k)$.

The aim now is to achieve $y(k) \approx r(k)$ by solving
an inverse problem dynamically. Replacing variables as in Sec.\,\ref{sub_comp_stat} and
omitting the noise term, (\ref{eq_model}) can be written as:
\begin{equation}
r(k) {=}  f^\ell\big(r(k-1), \ldots , r(k-n_y), m(k-\tau_{\rm d}), \ldots , m(k-n_u)\big). \nonumber
\end{equation}

As will become clear, it is convenient to introduce the time-shift $k \leftarrow k+\tau_{\rm d}$ --
meaning that $k$ will be replaced with $k+\tau_{\rm d}$ --
hence the last equation becomes:
\begin{eqnarray}
\label{eq_compensator_1}
r(k{+}\tau_{\rm d}) &{=}&  f^\ell\big(r(k{+}\tau_{\rm d}{-}1),\ldots,r(k{+}\tau_{\rm d}{-}n_y), \nonumber \\
					&~&	\hspace{1.4cm}		m(k),\ldots, m(k{+}\tau_{\rm d}{-}n_u)\big).
\end{eqnarray}

The aim is to find $m(k)$ that will drive the system to the desired target $r(k)$.
Consequently, as before, (\ref{eq_compensator_1}) will be expressed in terms of a
polynomial in the unknown $m(k)$ as:
\begin{eqnarray}
\label{eq_compensator_2}
0 {=} c_{{\ell_m}}(k)m(k)^{\ell_{m}} {+} c_{{\ell_m{-}1}}(k)m(k)^{\ell_m{-}1} +\ldots \!\!&\!\!+\!\!&\!\! c_{1}(k)m(k) \nonumber \\ 
   \!\!&\!\!+\!\!&\!\! c_{0}(k), 
\end{eqnarray}

\noindent
where the time-varying coefficients $c_{j}(k)$, $j = 0,\ldots,\ell_m$,  can depend on past values of $m$ up to time $k-1$,
and on past and future values of $r$ up to time $k+\tau_{\rm d}$. Hence, the following additional assumption
is required in the dynamical case.

\begin{assump}
	\label{Assump3}
	The reference signal must be known up to time $k+\tau_{\rm d}$.
\end{assump}

The following example illustrates this procedure.

\begin{example}
	\label{example4}
	Consider the same model used in Example\,\ref{ex_1}, replacing $y(k)$ with $r(k)$ and $u(k)$ with $m(k)$, yields:
	\begin{eqnarray}
	r(k) & {=} & \hat{\theta}_1r(k-1)+\hat{\theta}_2m(k-1)+\hat{\theta}_3m(k-1)m(k-2) \nonumber \\
	& {~} & + \hat{\theta}_4m(k-1)^2+\hat{\theta}_5m(k-1)^3, \nonumber
	\end{eqnarray}
	
	\noindent
	for which $\tau_{\rm d}=1$. Next, taking the time-shift $k \leftarrow k+1$, the last equation
	can be expressed in the form of (\ref{eq_compensator_2}):
	\begin{eqnarray}
	\label{mr1}
	\!\!\!\!\!\!0 &{=}& \hat{\theta}_5m(k)^3{+}\hat{\theta}_4m(k)^2{+}\big[\hat{\theta}_2{+}\hat{\theta}_3m(k{-}1)\big]m(k)\nonumber \\
	\!\!\!\!\!\!&{~}&{+}[\hat{\theta}_1r(k){-}r(k{+}1)], \nonumber \\
	\!\!\!\!\!\!0 &{=}& c_{3}(k)m(k)^3{+}c_{2}(k)m(k)^2{+}c_{1}(k)m(k)+c_{0}(k),
	\end{eqnarray}
	
	\noindent
	where all the values of $r$ are known (see Assumption\,\ref{Assump3}) and also all past values of $m$.
	Hence, at each time step $k$ the solutions to (\ref{mr1}), that is, the three values of $m(k)$ are the potential 
	compensation inputs. \hfill
	$\square$
\end{example}

In what follows, some important recommendations on initialization and the decision-making process adopted
to choose the root to be used as compensation input are stated.

\begin{remark}(Initial compensator conditions).
	\label{Rem1}
	From Assumption~\ref{Assump3}, $r(k+\tau_{\rm d})$ is known in compensator (\ref{eq_compensator_2}). Call that
	value $\bar{r}$. Using the calibration curve of ${\cal S}$ or the static nonlinearity of ${\cal M}$, find the
	respective $\bar{m}$ and take $m(j)=\bar{m},~j=k-1,\ldots,k{+}\tau_{\rm d}{-}n_u$. If the calibration curve is
	not available, this value can be obtained through the static nonlinearity of model ${\cal M}$ or by solving the
	static compensator $\bar{{\cal M}}_r$ (\ref{eq_comp_stat}), $\bar{r}{=}\bar{f}^{\ell}(\bar{m},\bar{r})$, for
	$\bar{m}$. If there is more than one solution
	to $\bar{r} =  \bar{f}^{\ell}(\bar{m},\bar{r})$ use the one that: i)~stabilizes the model output and ii)~satisfies
	constraints C\ref{c1} and C\ref{c2}. Items i) and ii) are automatically taken into account by using the calibration
	curve or static nonlinearity. \hfill $\square$
\end{remark}

\begin{remark}(The decision-making process).
	\label{Rem2}
	Let $\bm{m}_k \triangleq [m_{1}^k~\cdots~m_{\ell_m}^k]^T$ be the set of roots of (\ref{eq_compensator_2}). 
	If only one element of $\bm{m}_k$
	satisfies C\ref{c1} and C\ref{c2} , then this will be the compensation input at
	time $k$, otherwise we choose the appropriate root according to:
	\begin{alignat}{1}
	\label{c3}
	m(k)=\underset{m_j^k,~\forall j\in \{1,\ldots,\ell_m\}}{\arg\min}  & \Big(|m_j^k-m(k-1)|\Big) .   \\
	\text{subject to:}  ~C\ref{c1},\,C\ref{c2}\nonumber
	\end{alignat}
	
	\noindent
	The use of (\ref{c3}) selects the solution that is closest to the 
	compensation value used in the previous time step. This simple criterion results in smoother signals $m(k)$ 
	and, consequently, in less compensation effort \cite{Abreu_etal2020}. \hfill $\square$
\end{remark}

If $\ell_m$ is even and composed only of complex conjugate values, then take $m(k){=}m(k-1)$. 
This situation is not common for models that satisfy Assumption\,\ref{Assump1}. 
Algorithm\,\ref{sel_root} summarizes the method to select the appropriate root.

\begin{algorithm}
	\label{sel_root}
	\SetAlgoLined\SetArgSty{}
	\KwIn{$m(k-1); \quad \bm{m}_k \triangleq [m_{1}^k~\cdots~m_{\ell_m}^k]^T$}
	$v \leftarrow \infty$ \\
	$a \leftarrow 0$ \\
	\For{$j=1$ \KwTo $\ell_m$}{		
		\If{$m_j^k \in \mathbb{R}$ (C1) \textbf{and} $ u_{\min} \leq m_j^k \leq u_{\max}$ (C2)}
		{
			$e \leftarrow |m_j^k-m(k-1)|$\\
			\If{$e < v$}
			{	
				$v \leftarrow e$\\
				$a \leftarrow 1$\\
				$m(k) \leftarrow m_j^k$\\
			}
		}
	}
	\If{$a = 0$}
	{
		$m(k) \leftarrow m(k-1)$
	}	
	\KwOut{$m(k)$}
	\caption{Selecting the Appropriate Solution for \eqref{eq_compensator_2}}
\end{algorithm}

\vspace{-0.3cm}
\subsection{Compensation for Systems with Hysteresis}
\label{sub_comp_hys}

The inclusion of the first difference of the input $u(k)$ and the corresponding
sign function as regressors is a sufficient condition for NARX models to mimic
hysteresis loop \cite{Martins_Aguirre2016}. A general  NARX model set
\cite{bil_che/89} extended with these regressors 
will be referred to as ${\cal{M}}_{\rm h}$:
\begin{eqnarray}
\label{m_narx_hys}
\!\!\!\!\!y(k) &{=}& g^{\ell}\big(y({k-1}),\cdots,y(k-n_{y}), \,u(k-\tau_{\rm d}),\cdots, \nonumber \\ 
      &{~}& \hspace{3mm} u(k-n_u), \phi_{1}(k-1), \,\phi_{2}(k-1) \big) +e(k), 
\end{eqnarray}

\noindent
where $\phi_{1}(k){=}u(k){-}u(k{-}1)$, $\phi_{2}(k) {=} {\rm sign}(\phi_{1}(k))$, $g^{\ell}(\cdot)$
is a polynomial function of the regressor variables up to degree $\ell$, and the other parameters are
the same as defined in \eqref{eq_model}. For models such as \eqref{m_narx_hys}, there are two sets of
equilibria for the deterministic part (omitting the noise) under loading-unloading inputs: one for
loading with $\phi_{2}(k){=}1$, and one for unloading with $\phi_{2}(k){=}{-}1$ \cite{Abreu_etal2020}. A constrained approach is proposed and detailed by \cite{Abreu_etal2020} to ensure that the model can describe the dynamic behaviour and also features in steady-state. 

Therefore, to deal specifically with hysteresis compensation, the general compensation method proposed
in Sec.\,\ref{sub_comp_dyn} will be adapted. Considering non-constant inputs, the following simplification
will be used:
\begin{eqnarray}
\label{eq_product_simp}
\phi_1(k-1)\phi_2(k-1) & = & \phi_1(k-1)\dfrac{|\phi_1(k-1)|}{\phi_1(k-1)}, \nonumber \\
& = & |u(k-1)-u(k-2)|,
\end{eqnarray}

\noindent
for $u(k-1) \neq u(k-2)$.

The compensator is developed following the {\it steps} below:
\begin{enumerate}
	\item Rewrite ${\cal M}_{\rm h}$ as:\label{step_1}
	\begin{eqnarray}
	\label{mr_hys_gen_1}
	0 &{=}& g^{\ell}\big(y({k-1}),\cdots,y(k-n_{y}), \,u(k-\tau_{\rm d}),\cdots, \nonumber \\ 
	  &{~}&  u(k\!-\!n_u),\phi_{1}(k\!-\!1), \,\phi_{2}(k\!-\!1) \big) - y(k);
	\end{eqnarray}
	
	\item if $\phi_1(k-1)\phi_2(k-1)$ appears in any regressor of \eqref{mr_hys_gen_1}, use the result in \eqref{eq_product_simp}; \label{step_2}
	
	\item if $\phi_1(k\!-\!1)$ and $\phi_2(k\!-\!1)$ still appear, replace them with $u(k\!-\!1)\!-\!u(k\!-\!2)$ and 
	$|u(k\!-\!1)\!-\!u(k-2)|/[u(k\!-\!1)\!-\!u(k-2)]$, respectively; \label{step_3}
	
	\item if $[u(k-1)-u(k-2)]$ appears in any denominator, multiply the equation by $[u(k-1)-u(k-2)]$; \label{step_4}
	\item replace $y(k)$ with $r(k)$, $u(k)$ with $m(k)$, perform the time shift $k \leftarrow k + \tau_{\rm d}$, and
	rewrite this equation like (\ref{eq_compensator_2}); \label{step_5}
	\item split the equation with $|\cdot|$ into two polynomials in $m(k)$: \label{step_6}
	\begin{eqnarray}
	\label{mr_hys_load}
		0&{=}& c_{\ell_m}^{\rm L}(k)m(k)^{\ell_m}+c_{\ell_m-1}^{\rm L}(k)m(k)^{\ell_m-1}\nonumber \\
		&{~}&+\ldots+c_{1}^{\rm L}(k)m(k)+ c_{0}^{\rm L}(k) \\ 
		&{~}& \text{for} \quad  m(k)>m(k-1), \nonumber
	\end{eqnarray}
	\begin{eqnarray}
	\label{mr_hys_unload}
	0 &{=}& c_{\ell_m}^{\rm U}(k)m(k)^{\ell_m}+c_{\ell_m-1}^{\rm U}(k)m(k)^{\ell_m-1}\nonumber \\
	 &{~}&+\ldots+c_{1}^{\rm U}(k)m(k)+ c_{0}^{\rm U}(k)  \\ 
	  &{~}& \text{for} \quad  m(k)<m(k-1), \nonumber
	\end{eqnarray}
	
\end{enumerate}

\noindent 
where the superscripts $\rm L$ and $\rm U$ refer to loading and unloading regimes, respectively.
Note that, as we are now dealing with a hysteretic system, (\ref{eq_compensator_2}) has two
counterparts: one for loading  (\ref{mr_hys_load}), and one for unloading (\ref{mr_hys_unload}).
In this case, the compensation input $m(k)$ will be a feasible root of (\ref{mr_hys_load}) or (\ref{mr_hys_unload}).
Before detailing the decision-making process to compensate for such systems, the previous steps will be illustrated
below.

\begin{example}
	\label{example5}
	Suppose that the model ${\cal M}_{\rm h}$ is:
	\begin{eqnarray}
	\label{m_hys_1} 
	y(k) \!\!\!&\!\!\! = \!\!\!&\!\!\! \hat{\theta}_1y(k\!-\!1)\!+\! \hat{\theta}_2u(k\!-\!1)^3 
	\!+\! \hat{\theta}_3\phi_1(k\!-\!1)\phi_2(k\!-\!1)u(k\!-\!1) \nonumber \\ &  & +\hat{\theta}_4\phi_1(k-1)\phi_2(k-1)y(k-1). 
	\end{eqnarray}
	
	\noindent 
	Following steps\,\ref{step_1} and \ref{step_2}, \eqref{m_hys_1} is rewritten as:
	\begin{eqnarray}
	\label{m_hys_ex1}
	0 \!\!\!&\!\!\! = \!\!\!&\!\!\! \hat{\theta}_1y(k\!-\!1)\!+\! \hat{\theta}_2u(k\!-\!1)^3 
	\!+\! \hat{\theta}_3|u(k\!-\!1)-u(k\!-\!2)|u(k\!-\!1) \nonumber \\ &  & +\hat{\theta}_4|u(k\!-\!1)-u(k\!-\!2)|y(k-1) - y(k). 
	\end{eqnarray}
	
	\noindent
	In this model, steps\,\ref{step_3} and \ref{step_4} do not apply. By means of step\,\ref{step_5},
	we get the following equation (remember that $\tau_{\rm d}=1$):
	\begin{eqnarray}
	\label{m_hys_ex2}
	0 \!\!\!&\!\!\! = \!\!\!&\!\!\! \hat{\theta}_1r(k)\!+\! \hat{\theta}_2m(k)^3 
	\!+\! \hat{\theta}_3|m(k)-m(k\!-\!1)|m(k) \nonumber \\ &  & +\hat{\theta}_4|m(k)-m(k\!-\!1)|r(k) - r(k+1), 
	\end{eqnarray}
	
	\noindent
	which can be split into two polynomials (step\,\ref{step_6}) as:
	\begin{eqnarray}
	0 & = & \hat{\theta}_2m(k)^3 + \hat{\theta}_3m(k)^2+ \nonumber \\
	&~&[-\hat{\theta}_3m(k-1)+\hat{\theta}_4r(k)]m(k)+\nonumber \\
	&~& [\hat{\theta}_1r(k)-\hat{\theta}_4m(k-1)r(k)-r(k+1)] \nonumber \\
	&~& \text{for} \quad  m(k)>m(k-1), 
	\label{mr_hys_load_ex}
	\end{eqnarray}
	\begin{eqnarray}
	0 & = & \hat{\theta}_2m(k)^3 -\hat{\theta}_3m(k)^2+ \nonumber \\
	&~&[\hat{\theta}_3m(k-1)-\hat{\theta}_4r(k)]m(k)+\nonumber \\
	&~& [\hat{\theta}_1r(k)+\hat{\theta}_4m(k-1)r(k)-r(k+1)] \nonumber \\
	&~& \text{for} \quad  m(k)<m(k-1). 
	\label{mr_hys_unload_ex}
	\end{eqnarray}
	
	\noindent
	Here, \eqref{mr_hys_load_ex} refers to loading regime similar to form \eqref{mr_hys_load} with
	$c_{3}^{\rm L}(k) =\hat{\theta}_2$, $c_{2}^{\rm L}(k)=\hat{\theta}_3$,
	$c_{1}^{\rm L}(k)=-\hat{\theta}_3m(k-1)+\hat{\theta}_4r(k)$, and 
	$c_{0}^{\rm L}(k)=\hat{\theta}_1r(k)-\hat{\theta}_4m(k-1)r(k)-r(k+1)$; while, in an analogous way,
	\eqref{mr_hys_unload_ex} refers to unloading regime in form \eqref{mr_hys_unload} with coefficients 
	$c_{3}^{\rm U}(k) =\hat{\theta}_2$, $c_{2}^{\rm U}(k)=-\hat{\theta}_3$, 
	$c_{1}^{\rm U}(k)=\hat{\theta}_3m(k-1)-\hat{\theta}_4r(k)$, and 
	$c_{0}^{\rm U}(k)=\hat{\theta}_1r(k)+\hat{\theta}_4m(k-1)r(k)-r(k+1)$.

	The polynomials \eqref{mr_hys_load_ex} and \eqref{mr_hys_unload_ex}, which are
	valid for $k \geq 1$, can be initialized, at $k=0$, using an estimate of the hysteresis loop,
	as will be illustrated in Example~\ref{example6}. \hfill $\square$
\end{example}

In addition to Assumption\,\ref{Assump3},  C\ref{c1} and C\ref{c2}, the following must also
be true for hysteretic systems:
\begin{cons}
$m(k)>m(k-1)$, if \eqref{mr_hys_load} is used at time $k$; {\rm OR} 
\end{cons}

\begin{cons}
$m(k)<m(k-1)$, if \eqref{mr_hys_unload} is used at time $k$.
\end{cons}

Constraints C3 and C4 are needed to ensure that the root is consistent with the regime 
for which it was calculated. Therefore, the decision-making process for hysteretic systems
is similar to that explained in Remark\,\ref{Rem2} with the addition of these new constraints.
Hence:
\begin{alignat}{1}
\label{c4}
m(k)=\underset{m_{j}^k,~\forall j\in \{1,\ldots,\ell_m\}}{\arg\min}  & \Big(|m_{j}^k-m(k-1)|\Big),  \\
\text{subject to:}  ~C1,\,C2,\,Cq\nonumber
\end{alignat}

\noindent
where $q \in \{3,4\}$. The step-by-step procedure is analogous to Algorithm\,\ref{sel_root},
but now using \eqref{c4} instead of \eqref{c3}.

\subsection{Initialization of Compensators for Systems with Hysteresis}
\label{icsh}

If any parameter of compensators \eqref{mr_hys_load} and \eqref{mr_hys_unload} depends on
previous values of the compensation input $m(k)$, i.e. $\{m(k{-}1),\,\ldots,m(k{+}\tau_{\rm d}{-}n_u)\}$, 
such values must be determined for initialization. In Sec.\,\ref{sub_comp_dyn}, we can use the
static curve to estimate these initial values, as described in Remark\,\ref{Rem1}. Here,
a hysteresis loop ${\cal H}$ behavior displayed on the input-output plane will be used.
In what follows, a procedure is described to obtain ${\cal H}$ for a given
model ${\cal M}_{\rm h}$ (\ref{m_narx_hys}).

Consider the following sinusoidal input signal with period $T = 1/f_{\rm min}$:
\begin{equation}
\label{eq_u_quasi_static}
\tilde{u}(k) = A \sin\left(2\pi f_{\rm min}k\right)  + \tilde{u}_0, 
\end{equation}

\noindent 
where $f_{\rm min}=1/T$ is the lowest frequency of interest,
$\tilde{u}_0 {=} (u_{\rm min} {+} u_{\rm max})/2$ is the mean, and 
$A = u_{\rm max} - \tilde{u}_0$ is the amplitude. Using $\tilde{u}(k)$ in the model \eqref{m_narx_hys}, 
after the transient, the resulting data set $\{\tilde{u}(k), \tilde{y}(k)\}^{N_e}_{k=N_i}$, where $N_i > T$ and $N_e = N_i + T$,
correspond to a hysteresis loop $\cal{H}$. The use of  $\cal{H}$ to initialize the compensator is illustrated in the
next example.


\begin{example}
	\label{example6}
	Consider model\,(\ref{m_hys_1}) of Example\,\ref{example5}, whose parameter
	values are $\hat{\theta}_1=0.8$, $\hat{\theta}_2=0.4$, $\hat{\theta}_3=0.2$, and
	$\hat{\theta}_4 = 0.1$. In order to initialize the compensator, at $k=0$, $m(-1)$
	is needed in \eqref{mr_hys_load_ex} and \eqref{mr_hys_unload_ex}. For this purpose,
	suppose that the excitation input signal $\tilde{u}$ \eqref{eq_u_quasi_static}
	is defined with $A = 1$, $f_{\rm min} = 1 \text{ } {\rm Hz}$, and $\tilde{u}_0 = 1$ for
	which the resulting $\cal{H}$ is shown in Fig.\,\ref{fig_ini_hys_ex}.

	\begin{figure}[!h]
		\centering
		\includegraphics[width=0.8\columnwidth]{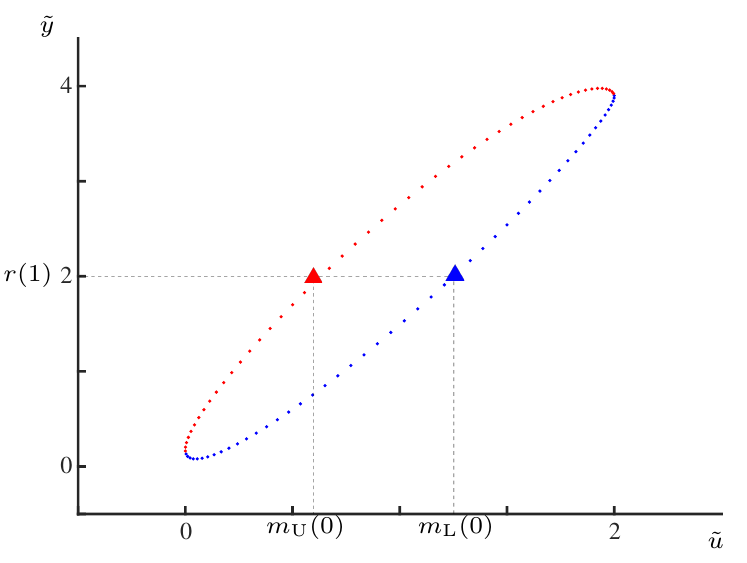}
	 	\vspace{-0.3cm}
		\caption{Loop $\cal H$ is obtained from the simulation of model \eqref{m_hys_1} with
			the input described in Example\,\ref{example6}. Blue dots (\textcolor{blue}{$\vdot$}) refer to loading regime,
			while those in red (\textcolor{red}{$\vdot$}) refer to unloading. For a given output or reference,
			say $r(1)$, there are two possible inputs indicated by triangles (\textcolor{blue}{$\blacktriangle$})
			blue and red (\textcolor{red}{$\blacktriangle$}) that correspond to $m_{\rm L}(0)$ and $m_{\rm U}(0)$,
			respectively. Which of these to use to start computing the compensator is determined by the regime at initialization time.}
		\label{fig_ini_hys_ex}
	\end{figure}

	From Assumption\,\ref{Assump3}, we have that the reference is known up to time $r(k{+}1)$.
	Suppose that $r(1)=2$ and that $\tilde{y}=r(1)$, there are two possible values for the input $\tilde{u}$,
	namely of, $ m_{\rm L}(0)$ and  $m_{\rm U}(0)$, which can be obtained from $\cal{H}$ (Fig.\,\ref{fig_ini_hys_ex}).
	The selection between these values is made based on the current regime of the reference signal, i.e., loading $(r(1)-r(0)>0)$
	or unloading $(r(1)-r(0)<0)$. In this example, at $k=0$, the reference is in the loading regime and, therefore, $m(0)=m_{\rm L}(0)$
	is chosen to initialize equations \eqref{mr_hys_load_ex} and \eqref{mr_hys_unload_ex}.  \hfill $\square$
	\vspace{-0.3cm}
\end{example}

\section{Results}
\label{NumericalExamples}

This section illustrates the compensator design
proposed in Sec.\,\ref{Methodology} for two simulated benchmark systems and for a pilot plant starting from the identified models.
The input design and other identification procedures are detailed in \cite{Tavares_2020_arxiv}. To evaluate the performance of the compensation achieved, the static nonlinearity of the compensated and uncompensated systems are compared and their time
evolution is evaluated using
the mean absolute percentage error ({\rm MAPE}) index is computed as follows:
\begin{equation}
	\label{Eq:MAPE}
	{\rm{MAPE}} = \dfrac{\sum_{k=1}^{N}|y_s(k) - y(k)|}{N|{\max}({\bm y_s})-{\min}({\bm y_s})|}.
\end{equation}

\vspace{-0.1cm}
\subsection{A Heating System}
\label{sec_hs_sys}

The bench test system is a small electrical heater modeled by the following
Hammerstein model \cite{agu_eal/05iee}:
\begin{align}
\label{heat_sys_s}
y(k)  = & \beta_1 y(k-1) + \beta_2 v(k-1) + \beta_3 y(k-2) + \beta_4 v(k-2), \nonumber \\ 
v(k)  = & p_1 u(k)^2+ p_2 u(k), 
\end{align}  

\noindent 
where $y(k)$ is the normalized temperature, and $u(k)$ is the electric power applied to the heater within the range
$ 0{ \leq} u(k) {\leq} 1$. The data set has been presented in \cite{agu_eal/02iee}, and is
available at \lucas{\textcolor{black}{\url{https://bit.ly/3iQ6rCF}}}.
The operation region of the model is $u(k) \in [0,~1]$ and $y(k) \in [0,~0.5]$. As described in \cite{Tavares_2020_arxiv}, the obtained parameters of (\ref{heat_sys_s}) are: $p_1 = 4.639331 \times 10^{-1}$, $p_2 = 5.435865 \times 10^{-2}$; while $\beta_1=1.205445$, $\beta_2 = 8.985133 \times 10^{-2}$, $\beta_3= -3.0877507 \times 10^{-1}$ and $\beta_4 = 9.462358  \times  10^{-3}$. From now on, the Hammerstein model (\ref{heat_sys_s}) will be treated as the system $\cal{S}$ to be compensated.

To compensate the nonlinearities in $\cal S$, the following three-term model $\cal M$ was obtained according to the procedure detailed in \cite{Tavares_2020_arxiv}:
\begin{equation}
\label{hs_model}
y(k) = \hat{\theta}_1y(k-1)+\hat{\theta}_2 u(k-2)^2 + \hat{\theta}_3y(k-2), 
\end{equation}

\noindent 
where $\hat{\theta}_1 = 8.958185 \times 10^{-1}$, $ \hat{\theta}_2 = 6.393347 \times 10^{-2}$, and
$\hat{\theta}_3 = -1.746750 \times 10^{-2}$. The validation results are shown in Table\,\ref{hs_tab_mape_model}, which indicate some degradation at higher frequencies and at points of
operation close to the origin.
\vspace{-0.2cm}
\begin{table}[htb]
	\centering
	\caption{${\rm MAPE}$ for model \eqref{hs_model} with sinusoidal inputs $u(k) = u_0 + 0.2 {\rm{sin}}(2\pi f k) $.
		Free-run simulation was used.}
	\setlength\tabcolsep{10pt} 
	\begin{tabular}{c | c c c}
		\multirow{2}{*}{$f$ [Hz]}
		& \multicolumn{3}{c}{$u_0$ [{\rm V}]} \\ 
		&  \textbf{$0.3$} & \textbf{$0.5$} & $0.7$ \\ \hline
		0.0005 & 5.5\% & 3.0\% & 2.9\%  \\  
		0.001 & 5.8\% & 2.9\% & 2.8\%  \\
		0.002 & 7.0\% & 4.0\% & 3.1\%
		
		\label{hs_tab_mape_model}	
	\end{tabular}
	\vspace{-0.1cm}
\end{table}

The static function of $\cal {M}$ (\ref{hs_model}) is: 
\begin{equation}
\label{hs_m_stab}
\bar{y} = \dfrac{\hat{\theta}_2 \bar{u}^2}{1-\hat{\theta}_1-\hat{\theta}_3},
\end{equation}

\noindent 
from where it is seen that for each value of the input, there is only one fixed point.
Because $\cal {M}$ is second-order, there are two eigenvalues at each fixed point $\bar{y}$.
The Jacobian matrix in this case does not depend on $\bar{u}$ or $\bar{y}$. 
Using (\ref{eq_stab_fix_point}) the condition for stability is:
\begin{equation}
	\begin{matrix}
		\left| \rm{eig} \left(
		\begin{bmatrix} 0 & 1 \\ \theta_3 & \theta_1 \end{bmatrix} 
		\right)
		\right|< 1, 
	\end{matrix} \nonumber
\end{equation}

\noindent 
where the eigenvalues of the Jacobian matrix are the algebraic solutions of
$\lambda^2-\theta_1\lambda-\theta_3 = 0$, which yields $|\lambda_1| = 0.8759$
and $|\lambda_2| = 0.0199$. Therefore, the fixed point for each input value is stable and, consequently, Assumption \ref{Assump2} is satisfied.

To illustrate the static compensation method presented in Sec.\,\ref{sub_comp_stat},
$\bar{y}$ is replaced with $\bar{r}$ and $\bar{u}$ with $\bar{m}$ in \eqref{hs_m_stab} to find
a polynomial in the unknown $\bar{m}$, that can be expressed like \eqref{eq_comp_stat}:
\begin{eqnarray}
\label{hs_bar_m}
\bar{r} & \!\!\! = \!\!\! &\dfrac{\hat{\theta}_2 \bar{m}^2}{1-\hat{\theta}_1-\hat{\theta}_3}, \nonumber \\
0 & \!\!\!=\!\!\! & \underbrace{[\hat{\theta}_2]}_{c_{m,2}}\bar{m}^2+\underbrace{[(\hat{\theta}_1+\hat{\theta}_3-1)\bar{r}]}_{c_{m,0}(\bar{r})}.
\end{eqnarray}

Since \eqref{hs_bar_m} is an incomplete quadratic equation and the operation region is limited to
$0 \leq \bar{m} \leq 1$, the algebraic solution is given by:
\begin{equation}
\label{hs_bar_m_2}
\bar{m} = \sqrt{\dfrac{-c_{m,0}(\bar{r})}{ c_{m,2}}} = \sqrt{\dfrac{(1-\hat{\theta}_1-\hat{\theta}_3)\bar{r}}{\hat{\theta}_2}}.
\end{equation}

\begin{figure}[htb]
	\centering
	\includegraphics[width=0.88\columnwidth]{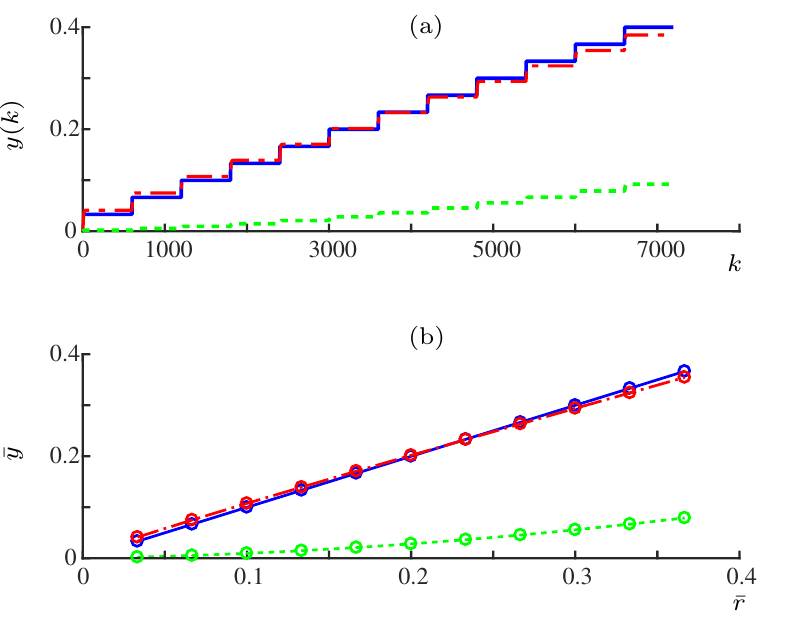}
	\normalsize \vspace{-0.3cm}
	\caption{Validation results for static compensation:
		(\textcolor{blue}{\hbox{--}}) is the reference; (\textcolor{red}{\hbox{-$\cdot$-}}), output of the compensated system; (\textcolor{green}{- -}), output of the uncompensated system. In (a), temporal evolution of the outputs and the reference; (b),  the 
		$r \times y$ plane.}
	\label{fig_hs_val_comp_stat_y}
	
	\vspace{-0cm}
\end{figure}

Also, as $\bar{r} \geq 0$, $\hat{\theta}_2 > 0$ and $1-\hat{\theta}_1-\hat{\theta}_3> 0$,
the result of the square root in \eqref{hs_bar_m_2} is always real. In Fig.\,\ref{fig_hs_val_comp_stat_y},
the static compensation results are shown for a reference that is a sequence of steps.
As expected,  the compensated system is approximately linear (see Fig. \ref{fig_hs_val_comp_stat_y}-b).
The static compensation can be used to find the initial values for the dynamical compensator when needed.

For dynamical compensation, using model $\cal M$ (\ref{hs_model}) 
 the procedure put forward in Sec.\,\ref{sub_comp_dyn} yields
(see Eq.\,\ref{eq_compensator_1}):
\begin{equation}
r(k+1) = \hat{\theta}_1 r(k)+\hat{\theta}_2m(k)^2+\hat{\theta}_3r(k-1) \nonumber
\end{equation}
	
\noindent
and (see Eq.\,\ref{eq_compensator_2}):
\begin{eqnarray}
\label{hs_mr_2}
\!\!\!\!0 & \!\!\!=\!\!\! & \underbrace{[\hat{\theta}_2]}_{c_{2}}m(k)^2+\underbrace{[\hat{\theta}_1r(k)+\hat{\theta}_3r(k-1)-r(k+1)]}_{c_{0}(k)}.
\end{eqnarray}

Solving \eqref{hs_mr_2} at each iteration yields the compensation input $m(k)$. Because
\eqref{hs_mr_2} is quadratic, there are two possibilities. Either both roots are real,
and then Algorithm\,\ref{sel_root} is used to select which one to use, or the roots are complex
conjugate in which case the previous value is used, i.e., $m(k)=m(k-1)$.

The algebraic solution of \eqref{hs_mr_2} gives the compensator $\cal M_{\rm r}$:
\begin{equation}
 m(k) = \sqrt{\dfrac{-c_{0}(k)}{c_{2}}} = \sqrt{\dfrac{r(k+1)-\hat{\theta}_1r(k)-\hat{\theta}_3r(k-1)}{\hat{\theta}_2}}. \label{hs_mr_3}
\end{equation}
\noindent
Because $m(k)$ does not depend on its previous values, in order to initialize \eqref{hs_mr_3}, 
it is sufficient to make $r(-1)=r(0)$ in (\ref{hs_mr_3}) at $k=0$. 

\begin{figure}[htb]
	\centering
	\includegraphics[width=0.88\columnwidth]{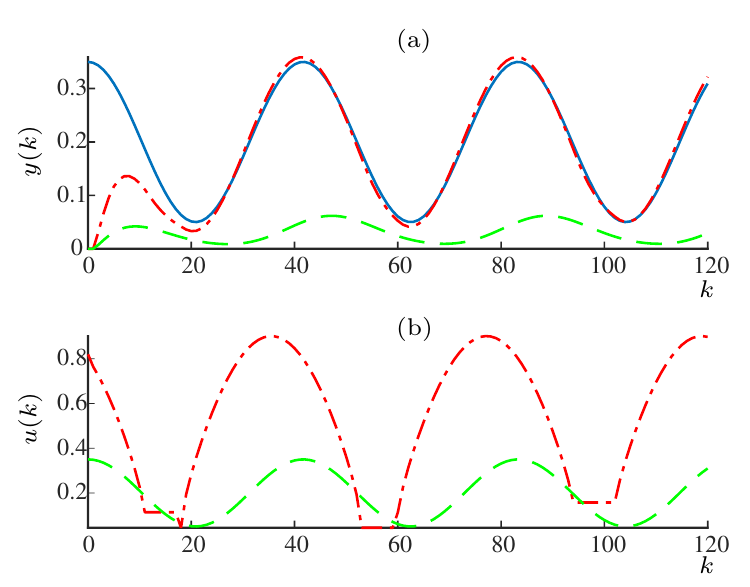}
	\normalsize \vspace{-0.3cm}
	\caption{Compensation results. (a)~temporal evolution; (b)~the applied inputs. In (a),
		(\textcolor{blue}{\hbox{--}}) is the reference $r(k) = 0.2 {\rm{sin}}(2\pi (0.002) k+\pi/2) + 0.15$; 
		(\textcolor{red}{\hbox{-$\cdot$-}}) is the output of the compensated system $y_{\rm c}(k)$ and
		(\textcolor{green}{\hbox{- -}}) is the output of the system $y_{\rm s}(k)$ without compensation. In (b), 
		(\textcolor{red}{\hbox{-$\cdot$-}}) is the compensation input $m(k)$, while (\textcolor{green}{- -}) is the input applied in the uncompensated system.}
	\label{fig_hs_val_comp_sin_y}
	\vspace{-0.3cm}
\end{figure}

In Fig.\,\ref{fig_hs_val_comp_sin_y} and in Table~\ref{hs_tab_mape_comp}, the results obtained with $\cal M_{\rm r}$ (\ref{hs_mr_3}) is compared to the uncompensated system for different reference signals. The uncompensated results are performed using $r(k)$ as the input for $\cal S$. From Table~\ref{hs_tab_mape_comp},
it is seen that as the frequency increases, the compensation becomes somewhat less effective, as would be expected for most control systems. Also for small values of $r(k)$ (see Fig. \ref{fig_hs_val_comp_sin_y}-b), complex roots appear, and according to Algorithm\,\ref{sel_root}, the last computed value will be used, that is, $m(k) = m(k-1)$. 

\begin{table}[htb]
	\centering
	\caption{${\rm MAPE}$ for  compensated and uncompensated systems with  $r(k) = r_0 {\rm{sin}}(2\pi f k+\pi/2) + r_0$.}
	\setlength\tabcolsep{10pt} 
	\begin{tabular}{c | c | c c c c}
		\multirow{2}{*}{} &\multirow{2}{*}{$f$ [Hz]}
		& \multicolumn{3}{c}{$r_0$ [{\rm V}]} \\ 
		& &  \textbf{$0.05$} & \textbf{$0.10$} & $0.20$ \\ \hline
		\multirow{4}{*}{Compensated} 
		& 0.0005 & 7.8\% & 4.1\% & 3.4\% \\  
		& 0.001 & 9.4\% & 6.4\% & 5.6\% \\
		& 0.002 & 15.5\% & 12.2\% & 10.2\% \\
		& 0.004 & 29.5\% & 25.8\% & 20.2\%
		\\ \hline
		
	\multirow{4}{*}{Uncompensated} 
	& 0.0005 & 45.6\% & 44.0\% & 40.8\% \\  
	& 0.001 & 45.5\% & 44.0\% & 40.9\% \\
	& 0.002 & 45.3\% & 44.0\% & 41.4\% \\
	& 0.004 & 44.8\% & 43.7\% &  \text{ }41.8\% 
		\label{hs_tab_mape_comp} 	
	\end{tabular}
\vspace{-0.6cm}
\end{table}

\begin{figure}[htb]
	\centering
	\includegraphics[width=0.8\columnwidth]{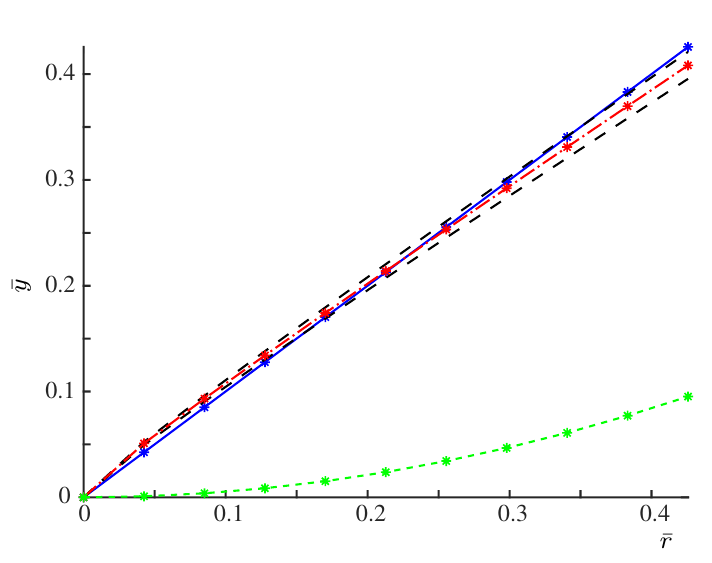}
	\vspace{-0.4cm}
	\caption{Results of a Monte Carlo test. 
	 (\textcolor{green}{-$\ast$-}) static curve of the uncompensated system, (\textcolor{blue}{-$\ast$-}) is the reference and 
	 (\textcolor{red}{-$\ast$-}) is $\mu(\bar{r})$, the mean calculated for the 10000 runs at each value $\bar{r}$, (\textcolor{black}{- -}) represent the error bars of $\pm 2\sigma(\bar{r})$.}
	
	\label{fig_hs_monte_carlo}
	\vspace{-0.8cm}
\end{figure}

Figure\,\ref{fig_hs_monte_carlo} shows the results for a Monte Carlo test of $10000$ runs.
During each run, a perturbed model $\cal M$ (\ref{hs_model}) is obtained by taking parameters
from a Gaussian distribution centered at the original parameters and with the covariance matrix
of the estimator. The black dashed lines indicate the region determined by $\mu(\bar{r}) \pm 2\sigma(\bar{r})$,
where $\mu(\bar{r})$ is the output mean and $\sigma(\bar{r})$ is the standard deviation.
The compensation performance is considerably effective on average for $\bar{r}{<}0.3$, and then it degrades a bit.

\subsection{A Hysteretic System}
\label{sec_bw}

In this example, the following Bouc-Wen model was used to describe the hysteretic behavior of a
piezoelectric actuator (PZT) that is an unimorph cantilever \cite{Rakotondrabe2011}:
\begin{align}
\dot{h}(t) =& \alpha_{\rm bw}\dot{u}(t)-\beta_{\rm bw}|\dot{u}(t)|h(t)-\gamma_{\rm bw} \dot{u}(t)|h(t)|,  \nonumber\\
y(t) =& \nu_{y}u(t) - h(t),
 \label{eq_bw}
\end{align}

\noindent
where $u(t)$[V] is the voltage input, $y(t)$[$\rm{\mu}m$] is the position output,
the parameters $\alpha_{\rm bw} = 0.9 [\mu\rm{ m}/\rm{V}]$ and
$\beta_{\rm bw}=\gamma_{\rm bw} = 0.008[\rm{V}^{-1}]$ determine the hysteresis loop, while
$\nu_{y}=1.6 [\mu\rm{ m}/\rm{V}]$ is a weight factor for the output. Here,
\eqref{eq_bw} is referred as the system $\cal S$ to be compensated, which is simulated
with a fourth-order Runge-Kutta method considering the integration step $\delta_t=5 \rm{ms}$.

The following NARX polynomial model ${\cal M}_{\rm h}$ (\ref{m_narx_hys}) to represent $\cal S$ was obtained as detailed in \cite{Tavares_2020_arxiv}:
\begin{eqnarray}
\label{bw_model}
y(k)\!\!\!\! &{=}&\!\!\!\!\hat{\theta}_1y(k{-}1){+}\hat{\theta}_2\phi_{2}(k{-}1)\phi_{3}(k{-}1)u(k{-}1) \nonumber \\
             & & {+}\hat{\theta}_3\phi_{2}(k{-}1)\phi_{3}(k{-}1)y(k{-}1){+}\hat{\theta}_4\phi_{2}(k{-}1), 
\end{eqnarray}

\noindent
where $\hat{\theta}_1 = 1.000099$, $ \hat{\theta}_2 = 6.630567 \times 10^{-3}$, $\hat{\theta}_3 = -6.247018 \times 10^{-3}$, and $\hat{\theta}_4=7.892915$. The validation results are shown in Table\,\ref{bw_tab_mape_model}, which indicate some degradation at higher frequencies and amplitudes.

\begin{table}[htb]
	\centering
	\caption{${\rm MAPE}$ for model (\ref{bw_model}) with sinusoidal inputs $u(k) =  {\rm{Gsin}}(2\pi f k)$.
		Free-run simulation was used.}
	\setlength\tabcolsep{10pt} 
	\begin{tabular}{c | c c c}
		\multirow{2}{*}{$f$ [Hz]}
		& \multicolumn{3}{c}{$\rm G$ [{\rm V}]} \\ 
		&  \textbf{$10$} & \textbf{$30$} & $50$ \\ \hline
		0.2 & 2.6\% & 2.0\% & 4.7\%  \\  
		1.0 & 2.7\% & 1.3\% & 4.1\%  \\
		5.0 & 7.7\% & 5.0\% & \text{ }3.6\% 
		\label{bw_tab_mape_model}	
	\end{tabular}
	\vspace{-0.2cm}
\end{table}

Following steps \ref{step_1}, \ref{step_2}, \ref{step_3} and \ref{step_5} for the
procedure presented in Sec.\,\ref{sub_comp_hys}, the compensator obtained is ${\cal M}_{\rm h, r}$ given by: 
\begin{eqnarray}
\label{hs_mr_hys_1}
0 \!\!&\!\!\!\! {=} \!\!\!\!&\!\! \hat{\theta}_1r(k){-}r(k{+}1){+}\hat{\theta}_2|m(k){-}m(k{-}1)|m(k) \nonumber \\
& &{+}\hat{\theta}_3|m(k){-}m(k{-}1)|r(k){+}\hat{\theta}_4|m(k){-}m(k{-}1)|,
\end{eqnarray}

\noindent
for which is assumed that $m(k) {\neq} m(k-1)$, and it can be split into two polynomials
in $m(k)$, like (\ref{mr_hys_load}) and (\ref{mr_hys_unload}) in step\,\ref{step_6}, as:
\begin{equation} 
\label{mr_hys_2}
\!\!0 {=} c_{2}^{\rm L}m(k)^2{+}c_{1}^{\rm L}(k)m(k){+} c_{0}^{\rm L}(k),~~ \text{for}~~  m(k){>}m(k{-}1);
\end{equation}
and
\begin{equation}
\label{mr_hys_3}
\!\!0{=} c_{2}^{\rm U}m(k)^2{+}c_{1}^{\rm U}(k)m(k){+} c_{0}^{\rm U}(k),~~ \text{for} ~~ m(k){<}m(k{-}1),
\end{equation}

\noindent
where $c_{2}^{\rm L}=\hat{\theta}_2$, $c_{1}^{\rm L}(k)=-\hat{\theta}_2m(k-1)+\hat{\theta}_3r(k)+\hat{\theta}_4$, $c_{0}^{\rm L}(k)=\hat{\theta}_1r(k)-\hat{\theta}_3m(k-1)r(k)-\hat{\theta}_4m(k-1)-r(k+1)$, $c_{2}^{\rm U}=-\hat{\theta}_2$, $c_{1}^{\rm U}(k)=\hat{\theta}_2m(k-1)-\hat{\theta}_3r(k)-\hat{\theta}_4$, and $c_{0}^{\rm U}(k)=\hat{\theta}_1r(k)+\hat{\theta}_3m(k-1)r(k)+\hat{\theta}_4m(k-1)-r(k+1).$

As some parameters of (\ref{mr_hys_2}) and (\ref{mr_hys_3}) depend on $m(k-1)$,
the initialization of the compensator is required at $k=0$. Applying
$\tilde{u} = 50{\rm{sin}}(2\pi0.2k)$ to model ${\cal M}_{\rm h}$ (\ref{bw_model}),
the loop ${\cal{H}}(\tilde{u}, \tilde{y})$ is determined. Making
$\tilde{y} = r(k + 1)$, $m(k-1)$ can be determined directly from
loop $\cal H$ similarly to Example\,\ref{example6}.

\begin{figure}[htb]
	\centering
	\includegraphics[width=0.88\columnwidth]{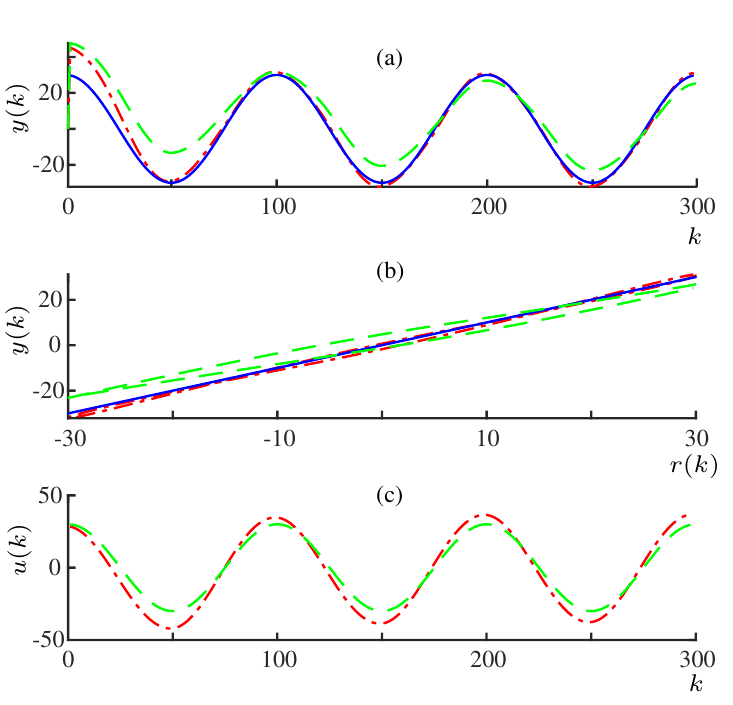}
	\vspace{-0.35cm}
	\caption{Compensation results for system \eqref{eq_bw}. (a)~temporal evolution of outputs; (b) the $r \times y$ plane and (c)~temporal evolution of inputs. (\textcolor{blue}{--}) represents the reference $r(k) = 30{\rm{sin}}(2\pi (2) k+\pi/2)$; 
		(\textcolor{red}{-$\cdot$-}), results for compensator given by (\ref{mr_hys_2}) and (\ref{mr_hys_3}) and (\textcolor{green}{- -}), uncompensated system.}
	\label{fig_bw_comp_valid}
\end{figure}

The validation results for compensation with (\ref{mr_hys_2}) and (\ref{mr_hys_3}) are shown in 
Fig.\,\ref{fig_bw_comp_valid} and in Table\,\ref{bw_tab_mape_comp}. These results
indicate that the compensated system presents better tracking performance than the 
uncompensated in all evaluated scenarios. In addition, the worst results occur
at higher frequencies and amplitudes.

\begin{table}[htb]
	\centering
	\caption{${\rm MAPE}$ for  compensated and uncompensated systems with  $r(k) = G_0 {\rm{sin}}(2\pi f k+\pi/2)$.}
	\setlength\tabcolsep{10pt} 
	\begin{tabular}{c | c | c c c c}
		\multirow{2}{*}{} &\multirow{2}{*}{$f$ [Hz]}
		& \multicolumn{3}{c}{$G_0$ [$\mu${\rm m}]} \\ 
		& &  \textbf{$20$} & \textbf{$30$} & $40$ \\ \hline
		\multirow{4}{*}{Compensated} 
		& 0.2 & 1.4\% & 3.2\% & 5.2\% \\  
		& 1.0 & 0.9\% & 2.5\% & 4.5\% \\
		& 2.0 & 1.0\% & 1.4\% & 3.4\% \\
		& 5.0 & 5.4\% & 4.5\% & 3.9\%
		\\ \hline
		
		\multirow{4}{*}{Uncompensated} 
		& 0.2 & 7.8\% & 7.1\% & 6.4\% \\  
		& 1.0 & 7.8\% & 7.0\% & 6.3\% \\
		& 2.0 & 7.7\% & 6.9\% & 6.1\% \\
		& 5.0 & 7.6\% & 6.6\% & \text{ }5.8\% 
		\label{bw_tab_mape_comp} 	
	\end{tabular}
\vspace{-0.2cm}
\end{table}

\begin{figure}[htb]
	\centering
	\includegraphics[width=0.88\columnwidth]{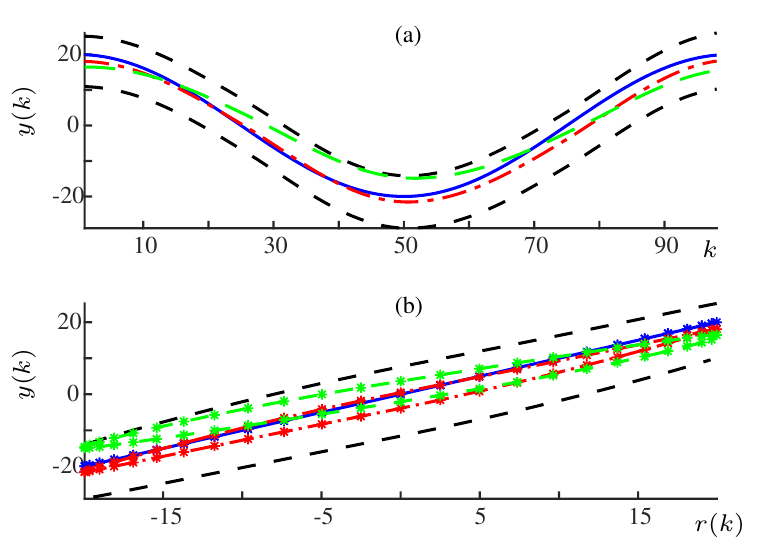}
	\vspace{-0.5cm}
	\caption{Results for 10000 Monte Carlo runs. (a) temporal evolution of the outputs in the last cycle; (b) plane $r(k) \times y(k)$.	(\textcolor{green}{-$\ast$-}) refers to the uncompensated system, (\textcolor{blue}{-$\ast$-}) refers to the reference and (\textcolor{red}{-$\ast$-}) is the average $\mu(k)$ of the 10000 Monte Carlo runs for the compensated system. (\textcolor{black}{- -}) represent the error bars of $\pm 2\sigma(k)$.}
	\label{fig_bw_monte_carlo}
	\vspace{-0.35cm}
\end{figure}

Finally, Fig.\,\ref{fig_bw_monte_carlo} shows the results for $10000$ Monte Carlo runs,
where $r(k) = 20 {\rm{sin}}(2\pi(2) k+\pi/2)$ is used in 5 cycles. A perturbed model ${\cal M}_{\rm h}$ (\ref{bw_model})
is obtained, as before. The Monte Carlo test presents a region determined by $\mu(k) \pm 2\sigma(k)$,
where $\mu(k)$ and $\sigma(k)$ are analogous to those described in Sec.~\ref{sec_hs_sys}, now for
each $k$, instead of $\bar{r}$. Comparing with Fig. \ref{fig_hs_monte_carlo}, the error bars determine a wider region. It is explained due to the sensitiviy of $\cal{M}_{\rm h}$ on variations in $\hat{\theta}_2$.

It should be mentioned that the constraint $\Sigma_y=1$ presented by \cite{Abreu_etal2020} is not fulfilled, where $\Sigma_y=1$ is the sum of all linear output's regressors. In order to show how this constraint affects the behavior of the model and compensator, we use a constrained least squares estimator to impose $\Sigma_y=1$ on the parameters of (\ref{bw_model}). The model obtained with constraints,
${\cal M}_{\rm h, cns}$, has the same structure of $\cal M_{\rm h}$ (\ref{bw_model}) with parameters: $\Sigma_y=\hat{\theta}_{1, \rm cns}= 1$, $\hat{\theta}_{2, \rm cns} = 6.630913 \times 10^{-3}$, $\hat{\theta}_{3, \rm cns} = -6.157515 \times 10^{-3}$, and $\hat{\rho}_{4, \rm cns}=7.893146$. 

Because the equal framework of ${\cal M}_{\rm h}$ and ${\cal M}_{\rm h, cns}$, their compensators
also have a same polynomial structure. The compensator  ${\cal M}_{\rm h, cns,r }$ is obtained when we replace each corresponding parameter of ${\cal M}_{\rm h, cns}$ in (\ref{mr_hys_2}) and (\ref{mr_hys_3}). Since the results obtained for modeling and compensation are similar to those presented by Tables\,\ref{bw_tab_mape_model} and \,\ref{bw_tab_mape_comp}, these results are omitted.

\begin{figure}[htb]
	\centering
	\includegraphics[width=1\columnwidth]{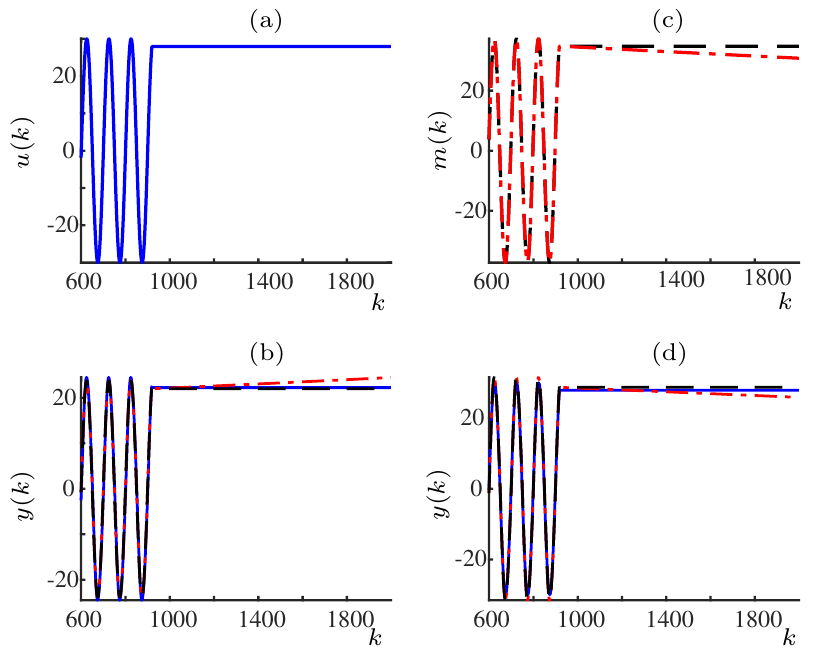}
	\vspace{-0.4cm}
	\caption{Validation and compensation results for models ${\cal M}_{\rm h}$ and ${\cal M}_{\rm h, cns}$  when a sinusoidal input or reference become constant. In (a), we have the temporal evolution of the input $u(k)=30{\rm sin}(2\pi(2)k)$ that becomes constant at $k=920$. This input determines the outputs in (b) when applied to the system $\cal S$ (\textcolor{blue}{\hbox{--}}), to ${\cal M}_{\rm h}$ (\textcolor{red}{\hbox{-$\cdot$-}}) and to ${\cal M}_{\rm h, cns}$ (\textcolor{black}{\hbox{- -}}). In (c), $m(k)$ for ${\cal M}_{\rm h}$ (\textcolor{red}{\hbox{-$\cdot$-}}) and ${\cal M}_{\rm h, cns}$ (\textcolor{black}{\hbox{- -}}) that were calculated for $r(k)=u(30{\rm sin}(2\pi(2)k)$) Finally, (d) shows the respective outputs for (c) where $r(k)$ is (\textcolor{blue}{\hbox{--}}).}
	\label{fig_bw_constant}
	\vspace{-0.3cm}
\end{figure}

A more relevant comparison for these two models and compensators is shown in Fig.\,\ref{fig_bw_constant}.
In this figure, the performance of the models and compensators is verified when an input/reference sinusoidal signal
becomes constant. Note that the validation and compensation results for ${\cal M}_{\rm h}$ (\ref{bw_model})
do not converge at steady-state. As $\bar{\phi}_1=\bar{\phi}_2=0$, both models become $\bar{y} = \Sigma_y\bar{y}$ that
have a single eigenvalue equal to $\Sigma_y$. Consequently, as $\Sigma_y=\hat{\theta}_1>1$, ${\cal M}_{\rm h}$ is unstable
in steady-state while the constraint $\Sigma_y=\hat{\theta}_{1,\rm cns}=1$, makes ${\cal M}_{\rm h, cns}$ remains in the last state. The compensation methods work in an open-loop, hence steady-state errors are expected, but they tend to be less
significant for more precise models.
For the current example, it was found that model ${\cal M}_{\rm h}$ has a steady-state error that tends to infinity over time, while the constrained model ${\cal M}_{\rm h, cns}$ provides a steady-state error of approximately 0,26$\mu$m in Fig. \ref{fig_bw_constant}-(b). Such errors are reflected in the
compensation performance since ${\cal M}_{\rm h, cns, r}$ presents an offset error of 0.83$\mu$m in Fig. \ref{fig_bw_constant}-(d) while the other, ${\cal M}_{\rm h, r}$, has a steady-state error which tends to infinity similarly to its model.

\subsection{Experimental Results}
\label{sec_exp_results}
\definecolor{cms}{HTML}{ff00ff}
\definecolor{cci}{HTML}{94a0a0}
\definecolor{inv}{HTML}{6cbbe7}

In this section, the compensation strategy is applied to an experimental pneumatic control valve,
which is a type of actuator widely used in industrial processes. For this type of actuator, the
control performance can degrade significantly due to 
friction, dead-zone, dead-band, and hysteresis \cite{Choudhury_etal2008,rom_gar/11}. 

The present valve is the same used in \cite{Abreu_etal2020}, where the measured output is its
stem position and the input is a pressure signal applied to the valve after passing V/I and I/P conversion.
The sampling time is $T_{\rm s}=0.01\,{\rm s}$ and, for details of the identification of this system the reader is
referred to \cite{Abreu_etal2020,Tavares_2020_arxiv}. The following models will be considered in this 
case study.

\vspace{0.2cm}
\noindent 1) ${\cal M}_{\rm h}$ is the model identified with the inclusion of $\phi_{1}(k)$ and $\phi_{2}(k)$ as
candidate regressors \cite{Martins_Aguirre2016}, and with the gray-box restrictions proposed by \cite{Abreu_etal2020}. As shown in Fig. \ref{fig_bw_constant}, the use of such constraints is important to describe the behavior in steady-state. The estimated model is
\begin{eqnarray}\label{mh_cns_valve}
y(k) & = & \hat{\theta}_1y(k-1)+\hat{\theta}_2y(k-2) + \hat{\theta}_3\phi_1(k-1) \nonumber \\
& + & \hat{\theta}_4u(k-1)\phi_{1}(k-1)\phi_{2}(k-1) \nonumber \\ 
& + & \hat{\theta}_5y(k-2)\phi_{1}(k-1)\phi_2(k-1),
\end{eqnarray}
with $\hat{\theta}_1=9.76\times 10^{-1}$, $\hat{\theta}_2=2.40 \times 10^{-2}$, $\hat{\theta}_3=1.19 \times 10^{-1}$,
$\hat{\theta}_4=3.76$ and $\hat{\theta}_5=-4.73$. Note that, $\Sigma_y = \hat{\theta}_1 + \hat{\theta}_2 = 1$.

The following models are found in the literature.

\vspace{0.2cm}	
\noindent 2) ${\cal M}_{\rm bw}$ is used to represent a BW model (\ref{eq_bw}). To estimate the valve output, its parameters were re-estimated using an evolutionary approach based on
niches, which is formulated in \cite{tavares2019}. These parameters are: $\alpha_{\rm bw} = 7.54 \times 10^{-1}$, $\alpha_{\rm bw}=-4.96$, $\gamma_{\rm bw}=- 3.61$ and $\nu_{y}=7.54 \times 10^{-1}$.

The last two models adopted were identified in \cite{Abreu_etal2020} for the same system under study and
with the same identification data. 

\vspace{0.2cm}
\noindent 3) ${\cal M}_{\rm h,2}$ was identified -- see Eq. 33 in \cite{Abreu_etal2020} -- with the same constraints used for
${\cal M}_{\rm h}$ (\ref{mh_cns_valve}), plus an additional one such that
the input signal can be isolated when writing the compensator equation.

\vspace{0.2cm}
\noindent 4) ${\breve{\mathcal M}}_{\rm h}$ was identified to describe the inverse relationship
between $u(k)$ and $y(k)$ of the valve -- see Eq. 34 in \cite{Abreu_etal2020}.
Therefore, the model provides $\hat{u}(k)$ given $y(k)$. The set of candidate regressors 
includes $\breve{\phi}_{1}(k)=y(k)-y(k-1)$ and $\breve{\phi}_{2}(k)={\rm sign}[\breve{\phi}_{1}(k)]$. 

The performance of the direct models, the first three, subject to sinusoidal inputs with different amplitudes are shown in
Table\,\ref{valv_tab_mape_valid}, which indicate that these models have similar efficiency by MAPE. Since ${\breve{\mathcal M}}_{\rm h}$ is an inverse model, which predicts the input signal instead of the
output and must be simulated from a smoothed version of $y(k)$ \cite{Abreu_etal2020}, we do not directly compare the MAPE accuracy of this model with the others. More details can be found in \cite{Tavares_2020_arxiv}.

\begin{table}[htb]
	\centering
	\caption{${\rm MAPE}$ for  models validation with sinusoidal inputs  $u(t) = G_0 {\rm{sin}}(2\pi (0.1) t+\pi/4)+3{\rm V}$.}
	\setlength\tabcolsep{10pt} 
	\begin{tabular}{c | c c c c c}
		\multirow{2}{*}{Model} 
		& \multicolumn{4}{c}{$G_0$ [V]} \\ 
		& \textbf{$0.45$} &\textbf{$0.55$} & \textbf{$0.65$} & $0.75$ \\ \hline
		
		\multirow{1}{*}{{\hspace{-0.4cm}1 - ${\cal M}_{\rm h}$}} 
		& 3.6\% & 3.0\% & 3.1\% & 4.9\% \\  
		
		\multirow{1}{*}{\hspace{-0.2cm}2 - ${\cal M}_{\rm bw}$} 
		& 3.9\% & 4.1\% & 4.5\% & 6.5\% \\  
		
		\multirow{1}{*}{\hspace{-0.15cm}3 - ${\cal M}_{\rm h, 2}$} 
		& 3.2\% & 3.5\% & 3.9\% & 5.7\% \\  
		
		\hline
		
	\end{tabular}\label{valv_tab_mape_valid}
\end{table}

For each model, the respective compensator is shown below. The first compensator is obtained following
the strategy presented in Sec. \ref{sub_comp_hys}.

\vspace{0.2cm}
\noindent
1) The compensator for ${\cal M}_{\rm h}$ (\ref{mh_cns_valve}) is ${\cal M}_{\rm h,r}$, obtained with the application of steps \ref{step_1}, \ref{step_2}, \ref{step_3}, \ref{step_5} and \ref{step_6}. ${\cal M}_{\rm h,r}$ is compound by two quadratic polynomials like (\ref{mr_hys_2}) and (\ref{mr_hys_3}) with parameters: $c_{2}^{\rm L}=\hat{\theta}_4$, $c_{1}^{\rm L}(k)=\hat{\theta}_3-\hat{\theta}_4m(k-1)+\hat{\theta}_5r(k-1)$, $c_{0}^{\rm L}(k)=\hat{\theta}_1r(k)+\hat{\theta}_2r(k-1)+\hat{\theta}_3m(k-1)-\hat{\theta}_5r(k-1)m(k-1)-r(k+1)$, $c_{2}^{\rm U}=-\hat{\theta}_4$, $c_{1}^{\rm L}(k)=-\hat{\theta}_3+\hat{\theta}_4m(k-1)-\hat{\theta}_5r(k-1)$ and $c_{0}^{\rm L}(k)=\hat{\theta}_1r(k)+\hat{\theta}_2r(k-1)-\hat{\theta}_3m(k-1)+\hat{\theta}_5r(k-1)m(k-1)-r(k+1)$. 

\vspace{0.2cm}
\noindent 2) The compensation law $m(t)$, for model ${\cal M}_{\rm bw}$, was proposed
by \cite{Rakotondrabe2011}, and is reffered as ${\cal M}_{\rm bw, r}$ given by:
\begin{equation}\label{valve_m_r_bw}
m(t)= \dfrac{10}{7.21}[r(t)+h(t)].
\end{equation}

\vspace{0.2cm}
\noindent 3) The compensator for ${\cal M}_{\rm h, 2}$ namely ${\cal M}_{\rm h, 2, r}$ is given by Eq. 35 in \cite{Abreu_etal2020}.

\vspace{0.2cm}
\noindent
4) Finally, also extracted from \cite{Abreu_etal2020}, ${\breve{\mathcal M}}_{\rm h,r}$, the compensator for  ${\breve{\mathcal M}}_{\rm h}$ is given by Eq. 36 in the mentioned paper.

\begin{figure}[htb]
	\centering
	\includegraphics[width=0.95\columnwidth]{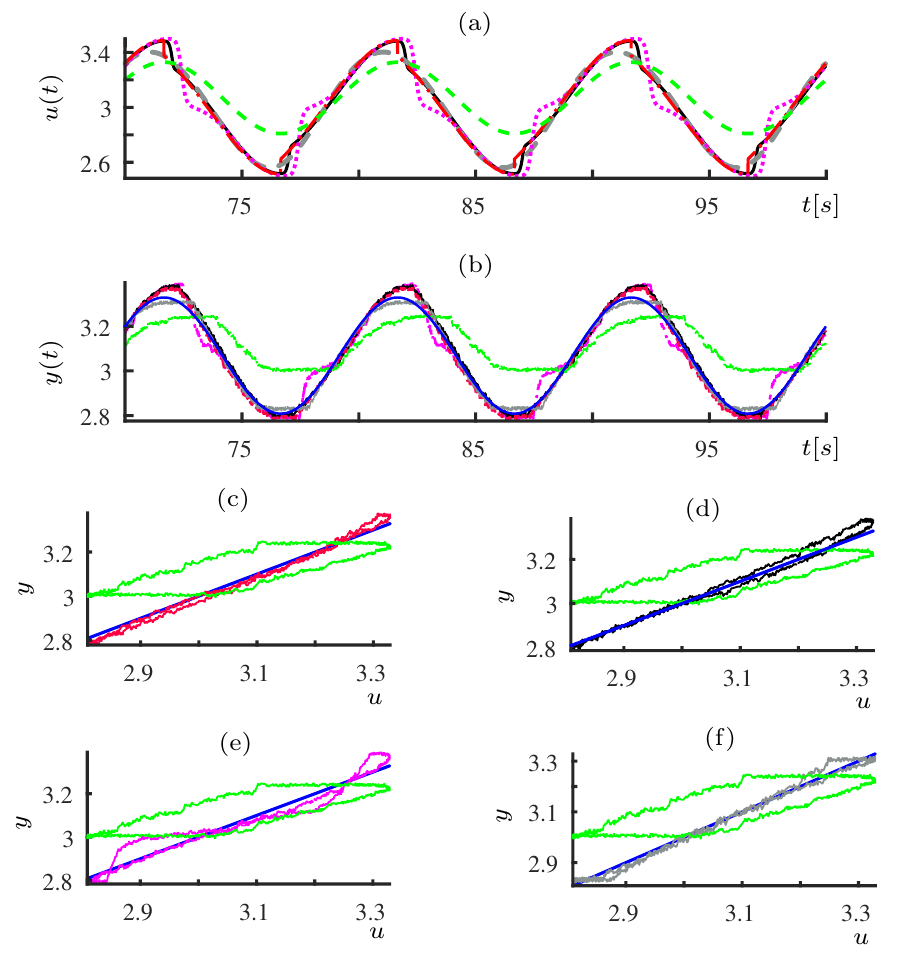}
	\vspace{-0.5cm}
	\caption{Compensation results for the pneumatic valve. (a)~temporal evolution of compensation inputs in three cycles; (b)~temporal evolution of outputs for the compensated systems in three cycles; (c)-(f) show the $r \times y$ plane for each compensator. (\textcolor{blue}{--}) refers to the reference $r(t) = 0.34{\rm{sin}}(2\pi (0.1)t+\pi/4)+3V$;  (\textcolor{green}{--}),  uncompensated system; (\textcolor{red}{-$\vdot$-}) compensation with  $\cal M_{\rm h, r}$\hbox{(\ref{mr_hys_2}-\ref{mr_hys_3})};  \hbox{(\textcolor{black}{--})} compensation with $\cal M_{\rm bw, r}$(\ref{valve_m_r_bw}); (\textcolor{cms}{$\vdot$ $\vdot$}) compensation with $\cal M_{\rm h, 2, r}$(35) in \cite{Abreu_etal2020}; (\textcolor{cci}{- -}) compensation with $\breve{\cal{M}}_{\rm h,r}$(36) in \cite{Abreu_etal2020}.}
	\label{fig_exp_comp_sin}
	\vspace{-0.3cm}
\end{figure}

The compensation results for sinusoidal references with different amplitudes are shown
in Table\,\ref{valv_tab_mape_comp} while Fig.\,\ref{fig_exp_comp_sin} shows the
compensation results for one of these references, $r(k)=0.41{\rm sin}(2\pi(0.1)t+\pi/4)+3{\rm V}$.
All compensation strategies provide considerably better results in all evaluated scenarios
when compared to the uncompensated system.

\begin{table}[htb]
	\centering
	\caption{${\rm MAPE}$ for  compensated and uncompensated systems with  $r(t) = G_0 {\rm{sin}}(2\pi (0.1) t)+3.07{\rm V}$.}
	\setlength\tabcolsep{6pt} 
	\begin{tabular}{c | c c c c c}
		\multirow{2}{*}{Compensation Strategy} 
		& \multicolumn{4}{c}{$G_0$ [V]} \\ 
		& \textbf{$0.26$} &\textbf{$0.34$} & \textbf{$0.41$} & $0.50$ \\ \hline
		
		\multirow{1}{*}{\hspace{-0.8cm} 1) $\cal M_{\rm h, r}$ (\ref{mr_hys_2}-\ref{mr_hys_3})}
		& 3.9\% & 3.3\% & 3.3\% & 3.8\% \\  
		
		\multirow{1}{*}{\hspace{-0.9cm}2) $\cal M_{\rm bw, r}$ (\ref{valve_m_r_bw})} 
		& 5.9\% & 4.4\% & 4.3\% & 4.2\% \\  
		
		\multirow{1}{*}{\hspace{0cm}3) $\cal M_{\rm h, 2, r}$ (35) in \cite{Abreu_etal2020}} 
		& 3.6\% & 3.5\% & 3.9\% & 5.2\% \\  
		
		\multirow{1}{*}{\hspace{-0.25cm}4) $\breve{\cal{M}}_{\rm h,r}$ (36) in \cite{Abreu_etal2020}} 
		& 2.9\% & 2.8\% & 3.2\% & 2.7\% \\ \hline
		
		\multirow{1}{*}{\hspace{-0.3cm}  Uncompensated} 
		& 21.0\% & 18.0\% & 16.2\% & 14.4\%  \\
		\hline

	\end{tabular}\label{valv_tab_mape_comp}
\end{table}

For the uncompensated system, the input is the reference $r(k)$. Using this as a
starting point, we would like to quantify how much more has to be done in order 
to achieve compensation. To this end, the following is computed
\begin{equation}
E({\delta_m}) = 
\sum_{k=N-N_0}^{N-1} \delta_m(k)^2,
\end{equation}

\noindent
where  $\delta_m(k)=|m(k)-r(k)|$ and $N$ is the length of $\delta_m(k)$.
$E({\delta_m})$ can be interpreted as the energy of $\delta_m(k)$ over one period $N_0$.
Also, the variability with respect to the uncompensated system is given by the
standard deviation of $\delta_m(k)$, $\sigma(\delta_m)$, see Table\,\ref{valv_tab_pot} for a summary of results.

 \begin{table}[htb]
	\centering
	\caption{$E({\delta_m})$[$\sigma({\delta_m})$] for the investigated compensators with  
	$r(t) = G_0 {\rm{sin}}(2\pi (0.1) t)+3.07{\rm V}$. }
	\setlength\tabcolsep{2.5pt} 
	\begin{tabular}{c | c c c c c}
		\multirow{2}{*}{Strategy} 
		& \multicolumn{4}{c}{$G_0$ [V]} \\ 
		& \textbf{$0.26$} &\textbf{$0.34$} & \textbf{$0.41$} & $0.50$ \\ \hline
		
		\multirow{1}{*}{\hspace{-0.7cm} 1) $\cal M_{\rm h, r}$ (\ref{mr_hys_2}-\ref{mr_hys_3})}
		& 28.3 [0.154]& 33.7 [0.165]& 39.4 [0.174]& 49.4 [0.186]\\  
		
		\multirow{1}{*}{\hspace{-0.8cm}2) $\cal M_{\rm bw, r}$ (\ref{valve_m_r_bw})} 
		& 27.0 [0.149]& 33.3 [0.163]& 40.6 [0.178]& 53.7 [0.197]\\  
		
		\multirow{1}{*}{\hspace{-0cm}3)$\cal M_{\rm h, 2, r}$ (35) in \cite{Abreu_etal2020}} 
		& 38.7 [0.184]& 43.7 [0.193]& 50.1 [0.203] & 57.2 [0.206]\\  
		
		\multirow{1}{*}{\hspace{-0.2cm}4)$\breve{\cal{M}}_{\rm h,r}$ (36) in \cite{Abreu_etal2020}} 
		& 24.9 [0.139]& 38.3 [0.174]& 52.7 [0.203]& 72.6 [0.237]\\ \hline

	\end{tabular}\label{valv_tab_pot}
\vspace{-0.3cm}
\end{table}

Similar to the validation results of model ${\cal M}_{\rm bw}$, the
corresponding compensator ${\cal M}_{\rm bw,r}$ (\ref{valve_m_r_bw}) performs slightly worse compared
to those based on NARX models. This suggests that NARX models are more appropriated to
describe nonlinearities in the valve. This is not surprising, as the class of NARX
polynomials is more general than the Bouc-Wen class.
On the positive side, the Bouc-Wen model provided the simplest compensator among those presented. In addition, 
as seen in Table\,\ref{valv_tab_pot} the respective compensator requires little change compared to the uncompensated
system. The most challenging task related to the Bouc-Wen model is to estimate its
parameters, which was done with an evolutionary approach.

Both compensation strategies proposed by \cite{Abreu_etal2020} performed well.
${\cal M}_{\rm h, 2, r}$ 
requires special care in the phase of structure selection otherwise the compensation input
$m(k)$ cannot be computed explicitly. Apart from that the compensation law tends to be easier 
to calculate than the one put forward in this work. On the negative side, ${\cal M}_{\rm h, 2, r}$
produces inputs with more abrupt changes (see Fig. \ref{fig_exp_comp_sin}-(a) and (e)). This
is reflected in higher values of $E(\delta_m)$ and ${\sigma}(\delta_m)$.

The overall good performance of compensator ${\breve{\mathcal M}}_{\rm h,r}$ comes at the 
expense of careful data preprocessing \cite{Abreu_etal2020}. 
This compensator can produce smooth compensation inputs (Fig.\,\ref{fig_exp_comp_sin}-(a) and (f)) with
low MAPE values (Table\,\ref{valv_tab_mape_comp}). However $E(\delta_m)$
and ${\sigma}(\delta_m)$ tend to increase considerably with the reference amplitude, as shown in Table \ref{valv_tab_pot}.

The compensator designed with the methodology put forward in this work, ${\cal M}_{\rm h, r}$, was
also able to compensate for the nonlinearity in the valve. The MAPE are among the lowest, especially for
moderate-high reference amplitudes, $G_0$, with the advantage that  $E(\delta_m)$ and ${\sigma}(\delta_m)$ 
do not increase as much as for the other regulators with $G_0$ (Table \ref{valv_tab_pot}). In addition, 
$\cal{M}_{\rm h}$ has only $5$ terms which facilitates obtaining the compensation law. Also, if the parameters
of such model needed to be updated, a recursive algorithm can be readily implemented. 
On the other hand, if the process models turn out to be polynomials with degree greater than 3, 
numerical solvers would be required to find the roots. Fortunately, many systems can be adequately
described using polynomials up to third-degree. As a side note, there is a self-consistency check indirectly
provided by the current method, which is the appearance of unfeasible roots: either real but outside the
operating range or complex. Whenever this happens it is an indication that the process model is not adequate.
Fortunately no such problems occurred in this case study.


\section{Conclusion}\label{Conclusion}

This work has presented an approach to compensate nonlinearities
based on NARX polynomial models previously estimated. The method
is simple and easy to interpret, as the compensation input turns out to
be the value required for the system to attain steady-state properties.
The compensation input is obtained iteratively, which confers some
adaptability to the method. The degree of adaptability can be readily 
increased by estimating the model parameters recursively, this has not
been explored in the paper. 

The method has been  considered 
in three contexts: static for constant references, dynamical for
variant, and for systems with hysteresis. At first, the technique was illustrated using two simulated systems.
The performance is comparable to that of other methods available in 
the literature.  In addition, the method presents some robustness to variation
in the parameters, as evaluated using Monte Carlo tests.

The proposed techinque was also implemented on a pilot plant where the goal was to 
compensate the nonlinearity of a pneumatic control valve. The performance 
was compared with a compensator designed in \cite{Rakotondrabe2011}
and two more recent strategies published by \cite{Abreu_etal2020}. 
All compensators can achieve nonlinearity compensation for the valve
(see Table\,\ref{valv_tab_mape_comp}). Pros and cons  of
each technique were discussed. 

Another interesting feature of the presented technique is that
a compensator can be designed for linear or nonlinear systems with or without using constraints
during model estimation. Perhaps the main foreseen limitation occurs if the compensators are 
designed using polynomial models of degree greater than $3$, which is not a common situation 
in practice, though it could happen. In this case, it would be necessary to use numerical solvers 
to find the roots, which could turn out to be a problem for more demanding online applications.
Fortunately, many relevant systems can be described by models up to 3rd
degree for which the roots can be found with analytical expressions presented in Appendix\,\ref{appendix_sol}.
 
Finally, the aim of the compensators is to cancel out most of the nonlinearity. This would allow for
the design of linear feedback controllers as a second step.


\appendices
\section{Solving Algebraic Polynomial Equations}\label{appendix_sol}
Algebraic polynomial equations with unknown $x$, degree $n \in \mathbb{N}^+$, and known coefficients $a_i \in \mathbb{R}$, $i \in \{0,\ldots,n\}$, can be expressed as
\begin{equation}\label{eq_poly}
0 = a_nx^n+a_{n-1}x^{n-1}+\ldots+a_1x+a_0, \quad a_n \neq 0
\end{equation} 
\noindent For \eqref{eq_poly}, there are $n$ complex roots. The analytical solutions for $n \leq 3$ are presented in the sequel.

\begin{itemize}
	\item[\textit{A)}] \textit{Linear Equations}
	\begin{align*}
		&0 = a_1x+a_0, \quad a_1 \neq 0 \nonumber \\
		&x = \dfrac{-a_0}{a_1} \nonumber
	\end{align*}

	\item[\textit{B)}]  \textit{Quadratic Equations}
	\begin{align*}
		&0 = a_2x^2+a_1x+a_0, \quad a_2 \neq 0 \nonumber \\
		&\Delta = a_1^2-4a_2a_0 \nonumber \\
		&x_i = \dfrac{-a_1 + (-1)^i \Delta}{2a_2}, \quad i \in \{0,1\}
	\end{align*}
	
	\item[\textit{C)}] \textit{Cubic Equations}
	\begin{align*}
		& 0  = a_3x^3+a_2x^2+a_1x+a_0, \quad a_3 \neq 0 \nonumber \\
		&\Delta_0 = a_2^2-3a_3a_1\nonumber \\
		&\Delta_1 = 2a_2^3-9a_3a_2a_1+27a_3^2a_0\nonumber \\
		&C = \sqrt[3]{\dfrac{\Delta_1 \pm \sqrt{\Delta_1^2-4\Delta_0^3}}{2}} \nonumber \\
		&\xi = \dfrac{-1+\sqrt{-3}}{2} \nonumber \\
		&x_i = -\dfrac{1}{3a}\left(b+\xi^iC+\dfrac{\Delta_0}{C}\right),  \quad i \in \{0,1,2\} \nonumber
	\end{align*}

\end{itemize}


\section*{Acknowledgment}

PEOGBA and LAA gratefully acknowledge financial support from CNPq
(Grant Nos. 142194/2017-4 and 303412/2019-4) and 
FAPEMIG (TEC-1217/98).	

\ifCLASSOPTIONcaptionsoff
  \newpage
\fi



\bibliographystyle{IEEEtran}
\bibliography{mychaos,idt_nls,reference}
\end{document}